\documentclass[longauth]{aa}  

\usepackage{txfonts}
\usepackage{graphicx}	
\usepackage{amsmath}	
\usepackage{textcomp, gensymb}    
\usepackage{xspace}
\usepackage{url}
\usepackage{natbib}
\bibpunct{(}{)}{;}{a}{}{,}
\usepackage[normalem]{ulem}

\makeatletter
\renewcommand*\aa@pageof{, page \thepage{} of \pageref*{LastPage}}
\makeatother

\newcommand{\mjysr}{MJy~sr$^{-1}$\xspace}
\usepackage[dvipsnames]{xcolor}
\definecolor{bblue}{RGB}{70, 130, 180}
\usepackage{svg}
\newcommand{\orcid}[1]{\href{https://orcid.org/#1}{\includegraphics[width=10pt]{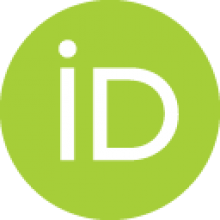}}}

\usepackage[colorlinks=true,linkcolor=bblue,citecolor=bblue,urlcolor=bblue]{hyperref}
\begin{document} 
   \title{JWST MIRI and NIRCam observations of NGC~891 and its circumgalactic medium}
   \author{
    {J\'er\'emy~Chastenet}\inst{\ref{UGent}}\orcid{0000-0002-5235-5589}\thanks{jeremy.chastenet@ugent.be} \and
    {Ilse~De~Looze}\inst{\ref{UGent}}\orcid{0000-0001-9419-6355}\thanks{PI: ilse.delooze@ugent.be} \and
    {Monica~Rela\~no}\inst{\ref{DeptGranada}, \ref{UGranada}}\orcid{0000-0003-1682-1148} \and
    {Daniel~A.~Dale}\inst{\ref{UWyoming}}\orcid{0000-0002-5782-9093} \and
    {Thomas~G.~Williams}\inst{\ref{Oxford}}\orcid{0000-0002-0786-7307} \and
    {Simone~Bianchi}\inst{\ref{INAF}}\orcid{0000-0002-9384-846X} \and
    {Emmanuel~M.~Xilouris}\inst{\ref{ObsAthens}}\orcid{0000-0002-2276-1412} \and
    {Maarten~Baes}\inst{\ref{UGent}}\orcid{0000-0002-3930-2757} \and
    {Alberto~D.~Bolatto}\inst{\ref{UMaryland}}\orcid{0000-0002-5480-5686} \and
    {Martha~L.~Boyer}\inst{\ref{STScI}}\orcid{0000-0003-4850-9589} \and
    {Viviana~Casasola}\inst{\ref{INAFRadio}}\orcid{0000-0002-3879-6038} \and
    {Christopher~J.~R.~Clark}\inst{\ref{ESAAURAatSTScI}}\orcid{0000-0001-7959-4902} \and
    {Filippo~Fraternali}\inst{\ref{Kapteyn}}\orcid{0000-0002-0447-3230} \and
    {Jacopo~Fritz}\inst{\ref{UMexico}}\orcid{0000-0002-7042-1965} \and
    {Fr\'ed\'eric~Galliano}\inst{\ref{CEA}}\orcid{0000-0002-4414-3367} \and
    {Simon~C.~O.~Glover}\inst{\ref{UHeidelbergAstro}}\orcid{0000-0001-6708-1317} \and
    {Karl~D.~Gordon}\inst{\ref{STScI},\ref{UGent}}\orcid{0000-0001-5340-6774} \and
    {Hiroyuki~Hirashita}\inst{\ref{ASIAA}, \ref{UOsaka}}\orcid{0000-0002-4189-8297} \and
    {Robert~Kennicutt}\inst{\ref{UArizona}}\orcid{0000-0001-5448-1821} \and
    {Kentaro~Nagamine}\inst{\ref{UOsaka}, \ref{UOsakaTJR},  \ref{KavliTokyo}, \ref{UNevada}}\orcid{0000-0001-7457-8487} \and
    {Florian~Kirchschlager}\inst{\ref{UGent}}\orcid{0000-0002-3036-0184} \and
    {Ralf~S.~Klessen}\inst{\ref{UHeidelbergAstro}, \ref{UHeidelbergInter}}\orcid{0000-0002-0560-3172} \and
    {Eric~W.~Koch}\inst{\ref{HarvardSmithsonian}}\orcid{0000-0001-9605-780X} \and
    {Rebecca~C.~Levy}\inst{\ref{UArizona}}\thanks{NSF Astronomy \& Astrophysics Postdoctoral Fellow}\orcid{0000-0003-2508-2586} \and
    {Lewis~McCallum}\inst{\ref{UStAndrews}} \and
    {Suzanne~C.~Madden}\inst{\ref{CEA}}\orcid{0000-0003-3229-2899} \and
    {Anna~F.~McLeod}\inst{\ref{UDurham}}\orcid{0000-0002-5456-523X} \and
    {Sharon~E.~Meidt}\inst{\ref{UGent}}\orcid{0000-0002-6118-4048} \and
    {Aleksandr~V.~Mosenkov}\inst{\ref{UBrigham}}\orcid{0000-0001-6079-7355} \and
    {Helena~M.~Richie}\inst{\ref{UPittsburg}}\orcid{0000-0001-6325-9317} \and
    {Am\'elie~Saintonge}\inst{\ref{UCollegeLondon}}\orcid{0000-0003-4357-3450} \and
    {Karin~M.~Sandstrom}\inst{\ref{UCSD}}\orcid{0000-0002-4378-8534} \and
    {Evan~E.~Schneider}\inst{\ref{UPittsburg}} \and
    {Evgenia~E.~Sivkova}\inst{\ref{UGent}} \and
    {J.~D.~T.~Smith}\inst{\ref{UToledo}}\orcid{0000-0003-1545-5078} \and
    {Matthew~W.~L.~Smith}\inst{\ref{UCardiff}}\orcid{0000-0002-3532-6970} \and
    {Arjen~van~der~Wel}\inst{\ref{UGent}}\orcid{0000-0002-5027-0135} \and
    {Stefanie~Walch}\inst{\ref{UKoln}} \and
    {Fabian~Walter}\inst{\ref{MPIA}}\orcid{0000-0003-4793-7880} \and
    {Kenneth~Wood}\inst{\ref{UStAndrews}}
}

\institute{
    Sterrenkundig Observatorium, Universiteit Gent, Krijgslaan 281-S9, 9000 Gent, Belgium
    \label{UGent} \and
    Dept. Fisica Teorica y del Cosmos, E-18071 Granada, Spain
    \label{DeptGranada} \and
    Instituto Universitario Carlos I de Fisica Teorica y Computacional, Universidad de Granada, E-18071 Granada, Spain
    \label{UGranada} \and
    Department of Physics and Astronomy, University of Wyoming, Laramie, WY 82071, USA
    \label{UWyoming} \and
    Sub-department of Astrophysics, Department of Physics, University of Oxford, Keble Road, Oxford OX1 3RH, UK
    \label{Oxford} \and
    INAF – Osservatorio Astrofisico di Arcetri, Largo E. Fermi 5, 50125 Firenze, Italy
    \label{INAF} \and
    Institute for Astronomy, Astrophysics, Space Applications \& Remote Sensing, National Observatory of Athens, P. Penteli, 15236 Athens, Greece
    \label{ObsAthens} \and
    Department of Astronomy and Joint Space-Science Institute, University of Maryland, College Park, MD 20742, USA
    \label{UMaryland} \and
    Space Telescope Science Institute, 3700 San Martin Drive, Baltimore, MD 21218, USA
    \label{STScI} \and
    INAF - Istituto di Radioastronomia, Via Piero Gobetti 101, 40129 Bologna, Italy
    \label{INAFRadio} \and
    AURA for the European Space Agency, STScI, 3700 San Martin Drive, Baltimore, MD 21218, USA
    \label{ESAAURAatSTScI}  \and
    Kapteyn Astronomical Institute, University of Groningen, P.O. Box 800, 9700AV Groningen, The Netherlands
    \label{Kapteyn} \and
    Instituto de Radioastronomía y Astrofísica, Universidad Nacional Autónoma de México, Morelia, Michoacán 58089, Mexico
    \label{UMexico} \and
    Université Paris-Saclay, Université Paris Cité, CEA, CNRS, AIM, 91191, Gif-sur-Yvette, France
    \label{CEA} \and
    Universit\"{a}t Heidelberg, Zentrum f\"{u}r Astronomie, Institut f\"{u}r Theoretische Astrophysik, Albert-Ueberle-Stra{\ss}e 2, D-69120 Heidelberg, Germany
    \label{UHeidelbergAstro} \and
    Institute of Astronomy and Astrophysics, Academia Sinica, Astronomy-Mathematics Building, No. 1, Section 4, Roosevelt Road, Taipei 10617, Taiwan
    \label{ASIAA} \and
    Theoretical Astrophysics, Department of Earth and Space Science, Osaka University, 1-1 Machikaneyama, Toyonaka, Osaka 560-0043, Japan
    \label{UOsaka} \and
    Theoretical Joint Research (TJR), Graduate School of Science, Osaka University, 1-1 Machikaneyama, Toyonaka, Osaka 560-0043, Japan
    \label{UOsakaTJR} \and
    Steward Observatory, University of Arizona, Tucson, AZ 85721, USA
    \label{UArizona} \and
    Kavli-IPMU (WPI), University of Tokyo, 5-1-5 Kashiwanoha, Kashiwa, Chiba, 277-8583, Japan
    \label{KavliTokyo} \and
    Department of Physics \& Astronomy, University of Nevada, Las Vegas, 4505 S. Maryland Pkwy, Las Vegas, NV 89154-4002, USA
    \label{UNevada} \and
    Universit\"{a}t Heidelberg, Interdisziplin\"{a}res Zentrum f\"{u}r Wissenschaftliches Rechnen, Im Neuenheimer Feld 205, D-69120 Heidelberg, Germany
    \label{UHeidelbergInter} \and
    Center for Astrophysics $|$ Harvard \& Smithsonian, 60 Garden Street, Cambridge, MA, 02138, USA
    \label{HarvardSmithsonian} \and
    School of Physics and Astronomy, University of St Andrews, North Haugh, St Andrews, KY16 9SS, UK
    \label{UStAndrews} \and
    Centre for Extragalactic Astronomy, Department of Physics, Durham University, South Road, Durham DH1 3LE, UK
    \label{UDurham} \and
    Department of Physics and Astronomy, N283 ESC, Brigham Young University, Provo, UT 84602, USA
    \label{UBrigham} \and
    Physics and Astronomy Department, University of Pittsburgh, 3941 O'Hara St, Pittsburgh, PA 15260
    \label{UPittsburg} \and
    Department of Physics \& Astronomy, University College London, London, UK, WC1E 6BT
    \label{UCollegeLondon} \and
    Department of Astronomy \& Astrophysics,  University of California, San Diego, 9500 Gilman Drive, La Jolla, CA 92093, USA
    \label{UCSD} \and
    Ritter Astrophysical Research Center, University of Toledo, Toledo, OH 43606, USA
    \label{UToledo} \and
    School of Physics \& Astronomy, Cardiff University, The Parade, Cardiff CF24 3AA, UK
    \label{UCardiff} \and
    I. Physikalisches Institut, Universi\"at zu K\"oln, Z\"ulpicher Str. 77, D-50937 K\"oln, Germany
    \label{UKoln} \and
    Max Planck Institute for Astronomy, K\"onigstuhl 17, D-69117 Heidelberg, Germany
    \label{MPIA}
}

   \date{Received X; X}

\abstract{
We present new JWST observations of the nearby, prototypical edge-on, spiral galaxy NGC~891.
The northern half of the disk was observed with NIRCam in its F150W and F277W filters. Absorption is clearly visible in the mid-plane of the F150W image, along with vertical dusty plumes that closely resemble the ones seen in the optical. 
A ${\sim }10 \times 3~{\rm kpc}^2$ area of the lower circumgalactic medium (CGM) was mapped with MIRI~F770W at 12~pc scales. Thanks to the sensitivity and resolution of JWST, we detect dust emission out to $\sim 4$~kpc from the disk, in the form of filaments, arcs, and super-bubbles. Some of these filaments can be traced back to regions with recent star formation activity, suggesting that feedback-driven galactic winds play an important role in regulating baryonic cycling.
The presence of dust at these altitudes raises questions about the transport mechanisms at play and suggests that small dust grains are able to survive for several tens of million years after having been ejected by galactic winds in the disk-halo interface. We lay out several scenarios that could explain this emission: dust grains may be shielded in the outer layers of cool dense clouds expelled from the galaxy disk, and/or the emission comes from the mixing layers around these cool clumps where material from the hot gas is able to cool down and mix with these cool cloudlets. 
This first set of data and upcoming spectroscopy will be very helpful to understand the survival of dust grains in energetic environments, and their contribution to recycling baryonic material in the mid-plane of galaxies.
}

   \keywords{Galaxies: individual: NGC~891 -- ISM: kinematics and dynamics  -- (ISM:) dust, extinction}

   \maketitle
%

\section{Introduction}
The disk-halo interface of galaxies is an active region, where inflow and outflow processes occur.
These mechanisms are fundamental to galaxy formation and evolution scenarios, as they strongly affect the star formation rate and metallicity of a galaxy, by the recycling of galactic material and inflows of pristine (``zero-metallicity'') gas. 
Understanding the interaction between the disk and the circumgalactic medium \citep[CGM; see a review by ][]{FaucherGigerePeng2023Review} is therefore important to constrain galaxy evolution, which is instrumental in comprehending the diversity of galaxies we observe today.

Gas in the disk-halo interface can come from gas accretion or cooling of the CGM, or ejected material from the mid-plane of galaxies \citep[][]{Marinacci2010, LehnerHowk2011, Marasco2022}.
In some cases, these processes can be tied to a single event, and take the form of galactic fountains \citep[][]{Bregman1980, Melioli2008, FraternaliBinney2006, Armillotta2016, Li2023}: after being expelled from the disk, material can cool down in the disk-halo interface and fall back down in the mid-plane of the galaxy.
The multitude of recent detections of massive galactic outflows in particular \citep{Veilleux2005, Veilleux2020} has reinforced their importance in regulating galaxy evolution, but their main driving forces (supernovae, stellar winds, ionising radiation, supermassive black holes, cosmic rays) have yet to be identified.
Characterising the efficiency of these various feedback mechanisms in regulating a galaxy’s vertical disk scale heights and in driving feedback-driven outflows are among the major challenges that current galaxy evolution models face \citep[e.g.,][]{Sancisi2008, FraternaliBinney2008, Marinacci2011, Gentile2013, NaabOstriker2017, Li2021, OtsukiHirashita2024}.

Simulations can be a powerful tool to understand these exchange mechanisms, and sort them with respect to their relative importance in driving gas in and out of the disk.
However, large-scale cosmological simulations typically do not resolve the extra-planar layers. Some discrepancies have been found between the mass outflow rates predicted by these simulations and the current observed rates.
On the other hand, high-resolution simulations can help focus on a direct comparison with disk-halo interface observations. Many works have focused on the gas distributions surrounding individual galaxies, through cosmological zoom-in simulations \citep{Kim2016AGORA, Marinacci2017, Hopkins2018, RameshNelson2024} and isolated galaxy simulations that self-consistently drive their outflows \citep{Fielding2017, Emerick2019, Martizzi2020, Schneider2020, Steinwandel2022}.
Additionally, tall-box simulations can capture the disk-halo connection in detail by simulating a small patch of the disk and a few kpc away from the mid-plane at particularly high resolution \citep{Walch2015, Girichidis2016, Gatto2017, KimOstriker2018, Suresh2019, Vijayan2020, Rathjen2021, Oku2022, McCallum2024, TanFielding2024}. These high resolution models tend to favor lower mass outflow rates more consistent with the observed values.

A more empirical approach is also often used. With kinematic measurements, several canonical works have studied the CGM of the Milky~Way and external galaxies \citep[][]{WakkervanWoerden1997, Fraternali2002, MarascoFraternali2011, Marasco2019}, with minimal impact from line-of-sight confusion. 
However, this effect becomes crucial when dealing with dust continuum emission. This is what makes observations of the CGM instinctively better in edge-on galaxies, offering the possibility to investigate the vertical extent of their different components and to draw a detailed picture of the vertical profiles of the stars, gas, and dust.
Several observations of the atomic gas have reported the presence of structures above the disk of galaxies \citep[][]{MatthewsWood2003, Boomsma2005, Zschaechner2015}, similar to ionised gas measurements \citep[][]{RossaDettmar2003, Heald2007, Kamphuis2008PhD}, and X-rays \citep[][]{Strickland2002, Strickland2004, HodgesKluckBregman2013}.
In this paper, we focus on the distribution of dust and stars in the disk-halo interface of the prototypical edge-on galaxy NGC~891 probed by the MIRI and NIRCam instruments on-board JWST. The unprecedented resolution and sensitivity of JWST let us resolve and study substructures in the CGM, which was never possible at these wavelengths.

A number of studies have focused on measuring the extent of the dust halo in other edge-on galaxies. 
Using the extinction of background quasars, \citet{Menard2010} found dust in the haloes of galaxies up to Mpc distances from their disks, revealing the presence of intergalactic dust \citep[although][nuance this result with \textit{Herschel} observations]{Smith2016}.
Several studies also use dust infrared (IR) emission, from mid- to the far-IR wavelengths, including the broad emission features at 3.3, 6.2, 7.7, 8.6, 11.3, 12.6, and 17~$\mu$m. These features are produced by vibrational and bending modes of bonds in carbonaceous dust (e.g., {C=C}, {C--H}). The population of grains responsible for these features is often referred to as Polycyclic Aromatic Hydrocarbons (PAHs), which are large aromatic molecules and a subclass of the more generic family of hydrogenated amorphous carbons \citep[e.g.,][]{Allamandola1985, Jones13, Li2020Review}. 
In NGC~5907, \citet{IrwinMadden2006} found that most of the mid-IR emission comes from the PAHs, rather than the very small grains, and that their emission extends further out than that of molecular gas, as far as 6.5~kpc, with a typical scale height of 3.5~kpc. This scale height is close to the one found in NGC~5529 by \citet{Irwin2007}, using ISO data. These authors also found that the PAH spatial distribution (namely in the vertical direction) in that galaxy correlates with H$\alpha$ emission. 
In NGC~5775, \citet{Lee2001} found a similar spatial correlation between ionised gas emission and the radio continuum.

Close-by and nearly perfectly edge-on ($d \sim 9.6~$Mpc, $i>89\degree$), NGC~891 is an ideal target to study the extent of a galaxy's extra-planar components outside of the main disk, and has been scrutinised as such. 
NGC~891 is classified as a normal star-forming spiral galaxy with similarities to the Milky~Way \citep{Swaters1997} but having a slightly higher global SFR.
The atomic hydrogen (\ion{H}{i}) in NGC~891 has been studied in several works \citep[][and references therein]{Swaters1997, Sancisi2008}. By analysing very deep interferometric observations, \citet{Oosterloo2007} have found that 30\% of the \ion{H}{i} is located in the extra-planar layer with features that are extending more than 20~kpc from the disk. The \ion{H}{i} extra-planar gas shows a slower rotation with respect to the gas in the disk, similar to other galaxies \citep[][]{Fraternali2002, Marasco2019}, a feature that is considered crucial to understand its origin \citep[e.g.,][]{Fraternali2017ASSL}.
The ionised gas content of the CGM of NGC~891 has also been the centre of numerous works. Besides the various structures visible in H$\alpha$ emission \citep[e.g.,][]{Rossa2004}, the works of, e.g., \citet{vdKruitSearle1981}, \citet{Dettmar1990}, and \citet{Rand1990} have heavily investigated the asymmetry of the ionised gas distribution, which was quickly attributed to the spiral arms of NGC~891. The velocity and structure of the ionised component were studied by \citet{Heald2006}, who found that the halo gas rotates more slowly than the disk gas, similarly to the atomic gas component \citep[see also][for velocity lag measurements in a sample of edge-on galaxies]{Levy2019}.
Later on, \citet{Kamphuis2007} used the attenuation in H$\alpha$ observations to infer the presence of dust out to $\sim 2$~kpc. 
Similarly, \citet{HowkSavage1997} studied the dust component seen in extinction, with work focused on conspicuous dust-rich structures expanding from the mid-plane out to the CGM \citep[see also work by, e.g.,][]{Alton2000}. 

Understanding the layered structure of spiral galaxies can be done in great detail in targets like NGC~891.
Studying the infrared emission of NGC~891 has thus taken a larger importance, from ground-based observations to the advent of infrared space-based facilities, observing the dust signature from mid- to far-IR and sub-millimetre wavelengths.
The work of \citet{Alton1998} provided the first deep imaging at 450 and 850~$\mu$m with the SCUBA instrument on the JCMT. 
Later, \citet{Popescu2004} used  170 and 200~$\mu$m photometry combined with optical, near- and mid-IR, and submm data to support a radiative transfer model of the galaxy \citep[][and following work]{Popescu2000}. 
Using multi-wavelength spectral energy distributions, \citet{Whaley2009} modelled the full IR emission of NGC~891. They found that PAHs extend further out than cold dust traced by sub-millimetre emission. 
Similar works have focused on reproducing empirical data with radiative transfer. Namely, \citet{BianchiXilouris2011}, modelling the \textit{Herschel}/SPIRE bands, found that the presence of dust vertically extends up to 12~kpc.
The emission from large grains was further studied by \citet{Bocchio2016} using \textit{Herschel} measurements. They found that the far-IR radial profiles in the PACS/SPIRE bands were better fitted using a two-layer disk. This was confirmed by \citet{Mosenkov2022} who used similar wavelengths, and additionally studied the correlation of stellar content with the dust distribution, in a sample of edge-on galaxies.
Note that the work of \citet{Schechtman-Rook2013} even suggested the existence of a ``super-thin disk'' of roughly $\sim 60 - 80~$pc in height to best reproduce near-IR data.
Complementary to these observations, the SOFIA aircraft surveyed the polarised emission of NGC~891 using the HAWC+ instrument. Studies by \citet{Jones2020} and \citet{Kim2023} focused on the orientation of the magnetic fields in the disk and found it was very closely aligned with and parallel to the mid-plane. The low polarisation fraction detected was suggested to be due to a highly turbulent medium, attributed to morphological components like spiral arms and bars (and possible instrumental uncertainties). Its extremely low value in the centre was hypothesised to be the signature of a bar, aligned with the line-of-sight \cite[see also ][]{GarciaBurilloGuelin1995}. 
Using near-IR polarimetry, \citet{Montgomery2014} revealed again a certain asymmetry, and used the deep polarisation data to compare with radio synchrotron emission. A similar analysis was done recently by \citet{Katsioli2023}, who focused on a wide multi-wavelength analysis to distinguish between dust, free-free, and synchrotron emission in and around NGC~891.
The advent of JWST can now probe the same layers seen in optical at comparable spatial resolution, without relying on larger beams like \textit{Herschel} instruments. With higher resolution, we will be able to eclipse beam smearing data and focus on detailed comparisons.

In this paper, we present the JWST Cycle~1 \hyperlink{https://www.stsci.edu/jwst/science-execution/program-information?id=2180}{{\tt GO~\#2180}} (PI: De Looze), focused on studying the cycle of baryonic material in NGC~891.
In Sect.~\ref{sec:data}, we outline the details of the acquisition of new JWST NIRCam \citep{Rieke2005NIRCam} and MIRI \citep{Rieke2015MIRI} observations and the data reduction pipeline we used. 
Section~\ref{sec:nircam} focuses on a qualitative description of the NIRCam images, with a brief look at the vertical stratification of the stellar emission.
Focusing on MIRI, we then analyse the distribution of the extraplanar grains in dusty plumes and clumpy structures, and report several newly identified superbubbles at several kpcs above the disk in Sect.~\ref{sec:miri}. 
The same section focuses on the implication of detecting 7.7~$\mu$m emission at high altitudes. We discuss the survival scenarios of small carbonaceous grains emitting at 7.7~$\mu$m above the disk, and how the detected extraplanar dust features connect back to the disk\footnote{In an upcoming paper, we will present MIRI~MRS spectroscopic observations of 4 positions at 0.5 and 1~kpc above the disk (each for two galactocentric radius positions), which will enable us to study the contribution of PAH emission to the MIRI broadband filters in more detail. Future work will also focus on characterising the clumpiness of the CGM material in NGC~891, and will present parallel multi-band MIRI imaging of several disk locations.}.
Section~\ref{sec:conclusion} summarises the conclusions of this paper.

\section{Data}
\label{sec:data}
In the following, we adopt a distance of 9.6~Mpc to NGC~891 \citep[][]{Strickland2004}, an inclination of 89.7\degree \citep{Xilouris1998, Xilouris1999} and a position angle of 22.9\degree \citep[][]{Bianchi2007, Hughes2014}.

\subsection{Observations}
\begin{figure*}
    \centering
    \includegraphics[width=\textwidth]{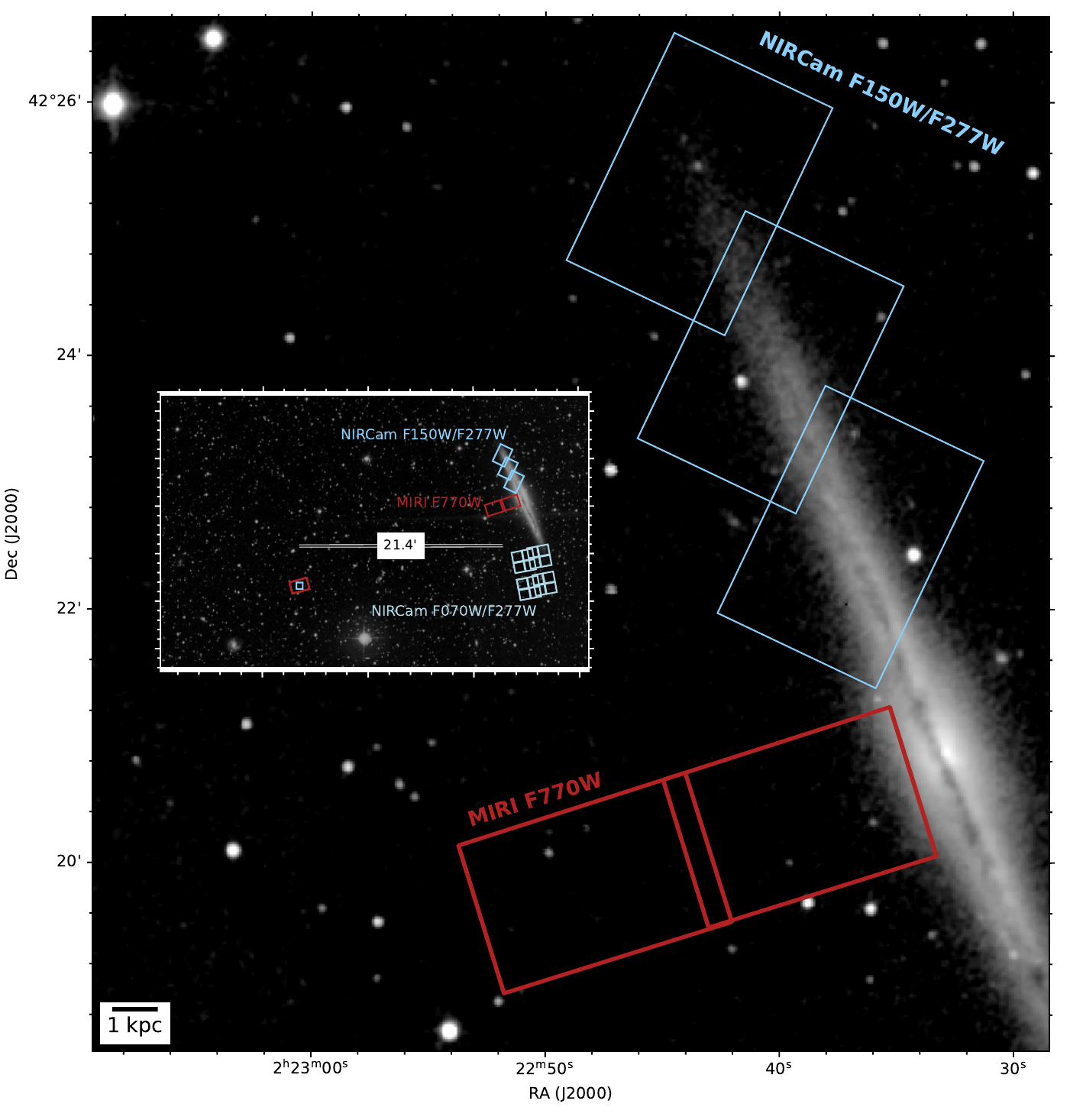}
    \caption{Large cutout of the 2MASS H-band image, showing the imaging patterns of the NIRCam F150W/F277W in blue, covering the north-half of NGC~891, and of the MIRI~F770W in red, covering $\sim 10$~kpc in length.
    The inset shows the positions of the dedicated background pointings, as well as the MIRI-parallel NIRCam F070W/F277W imaging (light blue), in the southern part of the galaxy, just outside of the disk.}
    \label{fig:propPatterns}
\end{figure*}

The JWST data used in this paper were taken as part of the Cycle~1 GO~\hyperlink{https://www.stsci.edu/jwst/science-execution/program-information?id=2180}{{\tt \#2180}} (PI: Ilse De Looze). 
NIRCam data using the short-wavelength F150W and long-wavelength F277W filters were taken on 15 August 2022. We targeted three fields in the northern half of the disk (selected as a compromise between overlap with ancillary data and time-feasibility), and obtained a mosaic from a 2-row--3-column pattern (10\% overlap). The observations were done in subarray {\sc sub}{\small 640} mode to reduce the risk of saturation in the disk, with a {\sc intramodulebox} dither strategy and 3 primary dithers, and 2 sub-pixel {\sc small-grid-dither}.
The MIRI~F770W observations were taken on December, 23$^{\rm rd}$ 2022.
The source pointing was centred on (02h22m43.5952s, +42\degree20$'$05.55$''$), and oriented to be perpendicular to the disk of the galaxy. 
The mosaic consists of two tiles (2-row--1-column, 10\% overlap), with a 4-point dither strategy. It covers vertical distance above the disk from $\sim 0.5$~kpc to $\sim 10$~kpc. 
In Fig.~\ref{fig:propPatterns}, we show the regions targeted by the NIRCam filters, in blue, focused on the northern half of the disk. The MIRI region is shown in red, covering mainly the CGM with limited emission from the disk.

For both NIRCam and MIRI, we obtained a background pointing centred on (02h24m39.1970s, +42\degree11$'$38.07$''$). This single-tile (not a mosaic) observation is located far away from the galaxy in an ``empty'' region of the sky, and is used to remove the astronomical background from the observations near/on the disk.
The inset in Fig.~\ref{fig:propPatterns} shows the separation between the source pointings and the background. The large distance between source and background pointings was motivated by the goal to detected faint CGM emission.
Note the light-blue pattern shows the parallel NIRCam F070W and F277W observations for reference, but the results of these parallel observations are not presented in this paper.

\subsection{Data reduction}
The level~3 MAST products suffer from significant artefacts that the default pipeline \citep{Bushouse2023} does not adequately handle, such as $1/f$ noise. 
Instead, we run {\sc PJPipe}, developed by the PHANGS-JWST team \citep[][]{Lee2023PHANGS}, and presented in \citet{williams2024}\footnote{The main difference in technical set-up between their sample and these observations is the that of NIRCam in sub-array mode. However, iterative improvements made {\sc PJPipe} effective for this set-up as well.}.
Further details of this pipeline can be found in \citet{williams2024}, describing all the steps with more information, as well as quality assurance figures for the PHANGS-JWST sample of galaxies.

Briefly, {\sc PJPipe} runs as follows, and unless mentioned otherwise, these steps apply to both MIRI and NIRCam: 
\vspace{-0.1cm}
\begin{itemize}
    \item stage~1: this step ({\tt calwebb\_detector1}) is kept to that of the default pipeline, described on the JWST pipeline website\footnote{\url{https://jwst-pipeline.readthedocs.io/en/latest/jwst/pipeline/}.}. Minor changes regarding the saturation flags are meant to attempt to recover more pixels than the default settings allow;
    \item first NIRCam destriping: this step removes the small-scale $1/f$ noise, well known to be present in the JWST NIRCam photometry. The large-scale structures are first filtered out, and the $1/f$ noise is removed using a row median in the vertical direction (first), then a median filter in the row direction;
    \item stage~2: this step ({\tt calwebb\_image2}) is also kept mostly to default settings as in the official pipeline. In this step, backgrounds get associated to the main source pointings and their effect can be adjusted by specific keywords;
    \item flux matching: the several dithers of our NIRCam data suffer from severe flux offsets. To alleviate these large offsets between dithers, the flux levels are matched between overlapping pixels and correction factors are found through matrix minimisation. This follows the approach of {\sc Montage} \citep[][]{Jacob2010}, but is done in two-steps: a ``dither-matching'', followed by a ``tile-matching'' on a stacked image, to increase the number statistics, due to the low overlap between dithers only;
    \item second NIRCam destriping: the large structure noise (initially filtered out) is still conspicuous in the NIRCam bands, especially in F150W. Using again medians across overlapping image, the large-structure stripes are removed, without affecting the galaxy emission;
    \item stage~3: the final step ({\tt calwebb\_image3}) combines the calibrated, de-noised exposure into a single mosaic, and performs the final adjustment of the background level.
\end{itemize}

In our case, the most outstanding remaining issues concern only F150W and the large-scale variations as well as offsets between some dither tiles. After several iterations of the pipeline and fine-tuning of several parameters, we find that it is impossible at this stage to perfectly correct for both. The image presented here represents the best compromise. 
Increasing the data quality would likely be done by increasing the number of dithers, to better average out the $1/f$ noise. It is possible that the {\sc sub-array} mode carries more subtleties (smaller array, quicker readout, etc).
For MIRI, only minimal artefacts remain. Some areas show negative values due to the offset-pointing obtained for the background subtraction. The background tile itself has bright sources that lead to over-subtraction in the main source pointings. We have tested several values for the parameters linked to that step, and found the best compromise, still showing a slight over-subtraction, which do not affect the conclusions of this paper.

\subsection{Extra-processing for NIRCam}
\label{sec:extraprocessing}
To limit the impact of the artefacts in the NIRCam images (uneven background, residual striping and dither offsets; Fig.~\ref{fig:finalimages}) on the preliminary results of this paper, we degrade our NIRCam images to a larger point spread function (PSF), and increase the pixel size.

We generate kernels for each NIRCam band to smooth to a 0\farcs2 2D-Gaussian using the {\sc WebbPSF} package, following the \citet{Aniano2011} procedure. The native PSFs are 0\farcs049 and 0\farcs088 for F150W and F277W, respectively. 
After convolution, we rebin the F277W data to $2\times$ bigger pixels, i.e. 0\farcs12 on a side, larger than the astrometric offset. 
The F150W image is projected to that same pixel grid after convolution as well.

Because of the evident gradient in the images, removing a single background value is not satisfactory.
Similarly, sigma-clipping the images and fitting a 2D plane to the remaining pixels proved unsuccessful, as the large artefacts strongly affect the outcome of this approach.
Instead, we select, by eye, a handful of small circular regions, where the flux distribution seems rather flat in F150W.
These regions are about $9''$ in radius so that they are much larger than the resolution element.  We fit a 2D plane to these regions only, so that the biggest (dither-related) variations are not taken into account, and only the broader gradient is fit. We then subtract this plane from the main image. The same ``background'' removal is done to the F277W image.
We note that this approach still provides limited-quality results, and more elaborate improvements are necessary in the future to correct the stripes and dither offsets within the pipeline. These will be shared along with any post-processing (background removal) once more science-ready images have been produced.

\subsection{Final images}
\begin{figure*}
    \centering
        \includegraphics[width=\textwidth, clip, trim={0 2cm 0 3cm}]{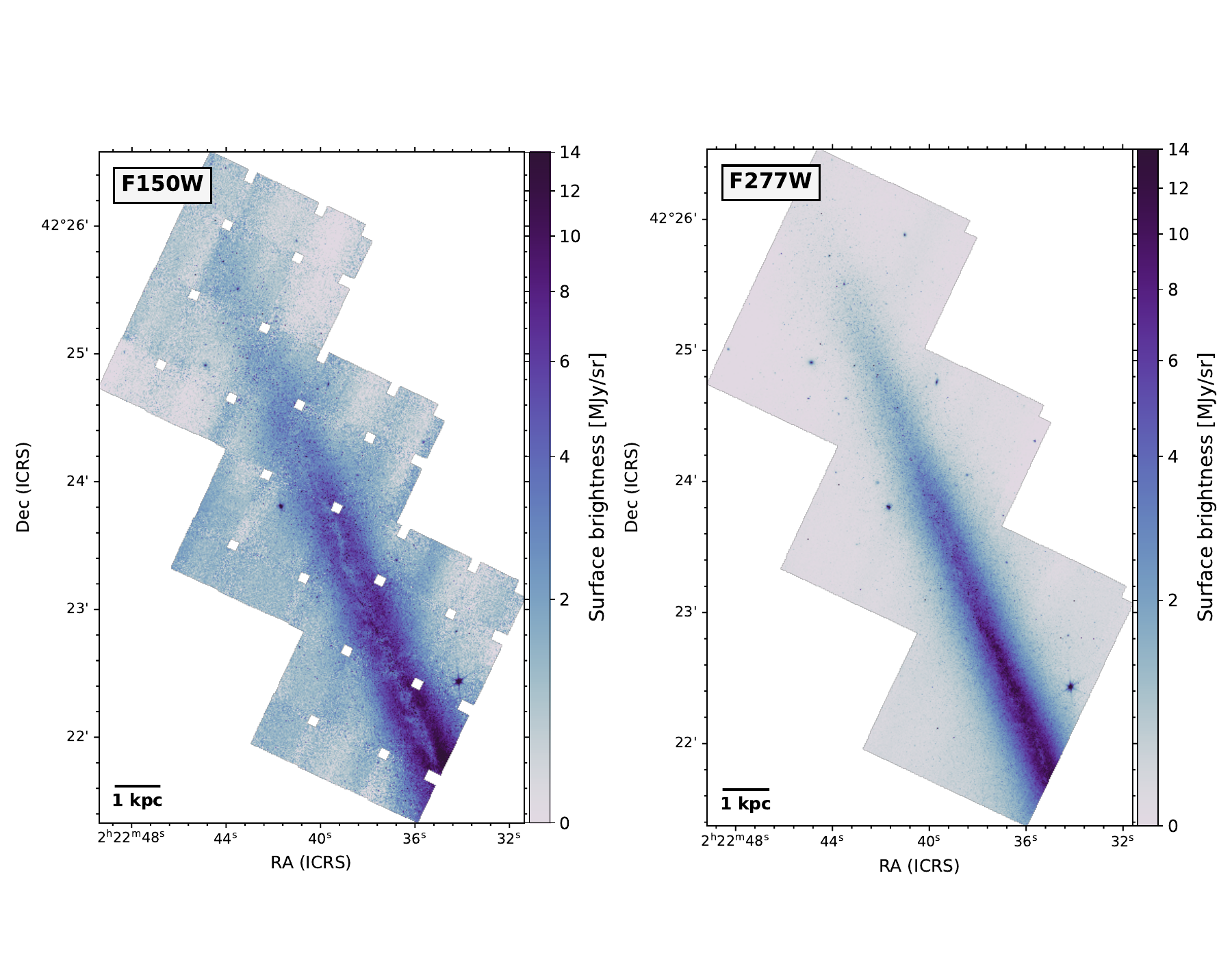}
        \includegraphics[width=\textwidth]{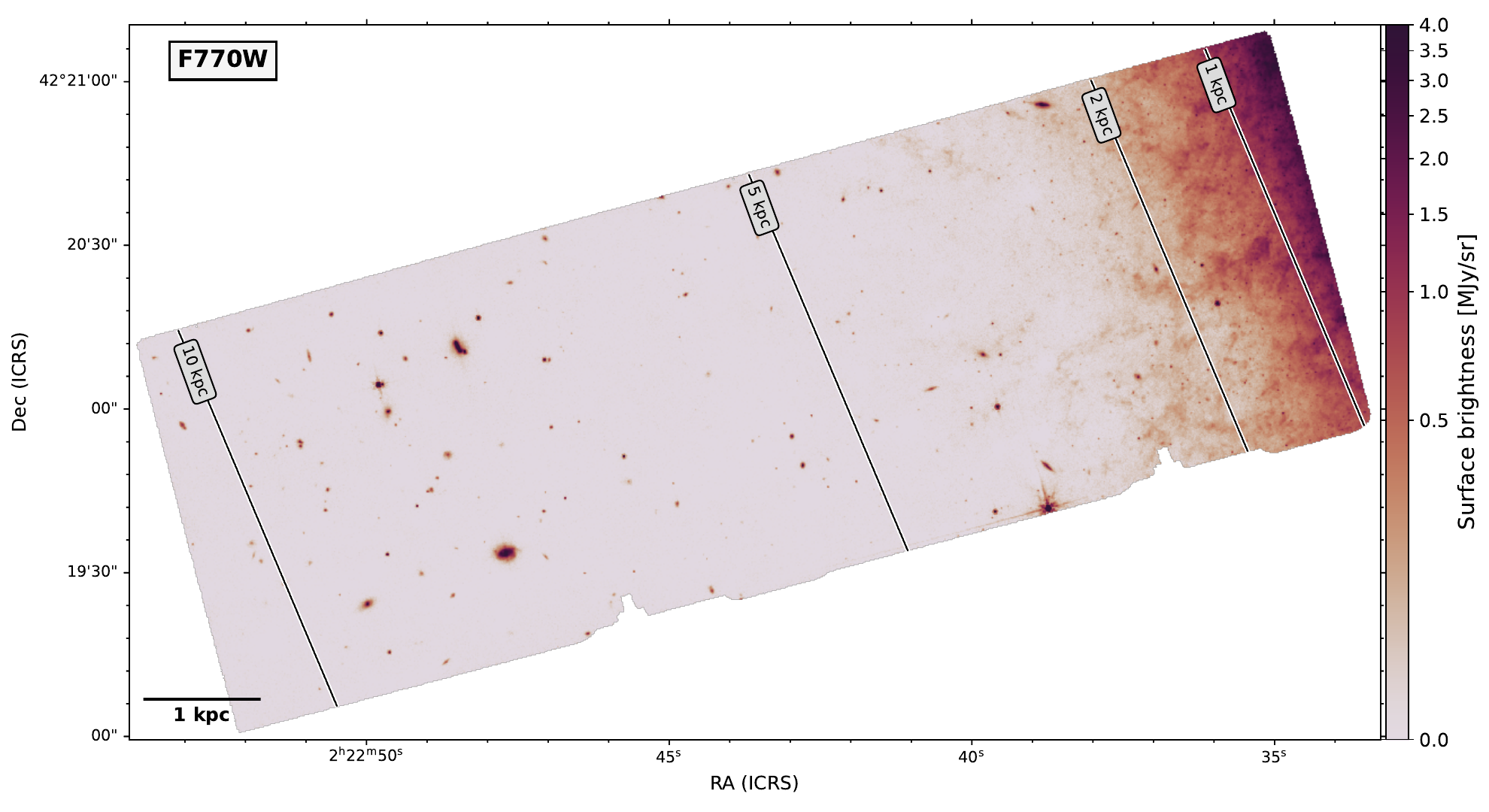}
    \caption{\textit{Top left:} NIRCam F150W data, at native $\sim 0.05''$ resolution. The mid-plane shows conspicuous absorption. The background in that band shows large discrepancies, due to the difficulty to correct for both large structure stripes and offsets between the individual dithers. The small squares are gaps in the dither-pattern coverage.
    \textit{Top right:} NIRCam F277W data, at native $\sim 0.092''$ resolution. Some absorption is visible in the mid-plane, and the image shows a much smoother background than for the F150W mosaic. 
    \textit{Bottom:} MIRI F770W data, at native $\sim 0.25''$ resolution, with radial distance indicators at 1, 2, 5, and 10~kpc from the mid-plane. The mid-IR image shows many filamentary structures as far out as $\sim 4$~kpc.}
    \label{fig:finalimages}
\end{figure*}

In Fig.~\ref{fig:finalimages}, we show the NIRCam~F150W and F277W, and the MIRI~F770W final images at their native resolution $\sim 0.05''$, $\sim 0.092''$, and $\sim 0.25''$, respectively.
We briefly and qualitatively describe these images.

\paragraph{NIRCam ---}
The northern half of the stellar disk is well resolved in both the F150W and F277W images. In the NIRCam~F227W image, the background is smooth, and we detect several foreground stars and background galaxies. Dust extinction effects are most prominently visible in the NIRCam~F150W image, whereas highly obscured regions also remain opaque to stellar radiation in the F277W image. We will discuss the NIRCam images in more detail in Sect.~\ref{sec:nircam}.
The small square gaps in the F150W image result from the dither pattern of the observations.

\begin{figure*}
    \centering
        \includegraphics[clip, trim={0 3cm 0 4cm}, width=\textwidth]{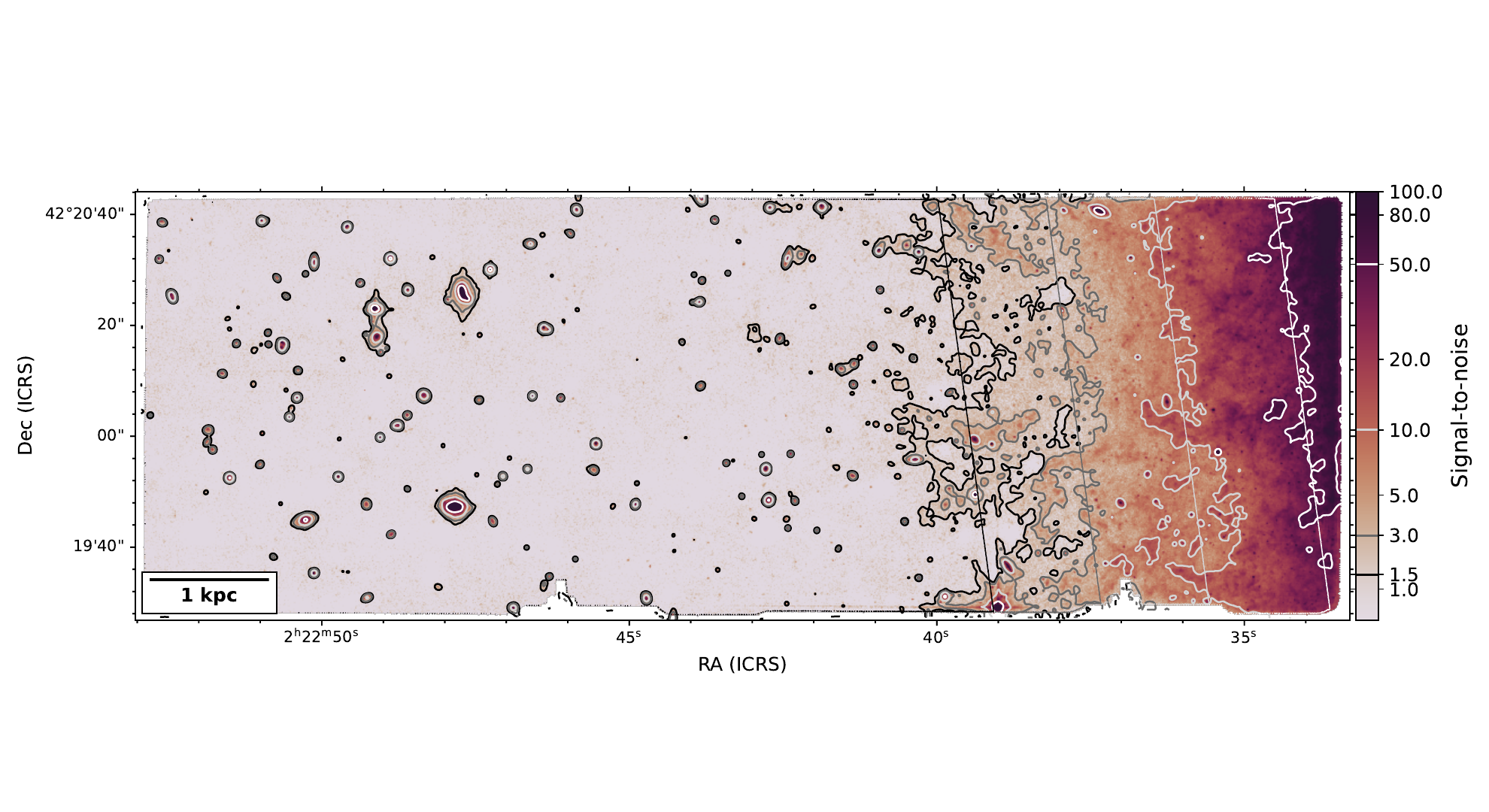}
    \caption{Signal-to-noise (S/N) map of the MIRI~F770W map, created using the error-map provided by the pipeline. The black, dark-grey, light-grey, and white contours show the S/N thresholds at 1.5, 3, 10, and 50 (with straight lines marking the associated distances at 3.8, 2.9, 2, 1~kpc), respectively, and are marked in the colorbar. These thresholds fall at distances of 3.8, 2.9, 2, and 1~kpc from the mid-plane.}
    \label{fig:miri_snr}
\end{figure*}

\paragraph{MIRI F770W ---}
The MIRI F770W image mosaic shows several high-resolution filaments close to the disk. In Fig.~\ref{fig:finalimages}, we overlay vertical distance indicators at 1, 2, 5, and 10~kpc.
There are also several bright sources far out from the disk, which can be either identified known sources (based on a query of the {\sc Simbad}\footnote{\url{http://simbad.cds.unistra.fr/simbad/}} database), or potentially high-redshift galaxies (not the focus of this paper). 
Fig.~\ref{fig:miri_snr} shows a signal-to-noise (S/N) map for the MIRI strip. We added contours at ${\rm S/N = 1.5, 3, 10, 50}$, which land roughly at 3.8, 2.9, 2, and 1~kpc, respectively. This map is created using the error-map provided by the pipeline (see \citealt{williams2024} for additional tests on error propagation).
Note that these S/N threshold positions will evolve as the MIRI image gets refined with pipeline improvements, and do not affect the qualitative analyses in this paper.
Using an iterative 3$\sigma$ clipping on the whole map, after masking over-subtracted sources, we find a noise median of $\sim 0.0036$~\mjysr and a large standard deviation of $\sim 0.017$~\mjysr. 
After comparing with the pipeline-produced error map, we can confirm that the large standard deviation is due to instrumental noise rather than related to astronomical variations in the background.

\paragraph{Saturated sources ---}
Some sources appear severely saturated in these images: in the NIRCam images, one is located in the bottom-right and two very close stars towards the mosaic centre, and in the MIRI image, there is one just at the lower-edge of the tile.
In the rest of the analysis, we mask these bright, saturated sources. We use the {\sc WebbPSF} package \citep{Perrin2014WebbPSF} to create PSF masks and manually align them with these sources.
Other bright sources do not show notable saturation spikes, and a more elaborate PSF masking will be developed in upcoming papers.

\subsection{Ancillary Data}
\paragraph{Other mid-IR data ---} We use the IRAC~8~$\mu$m and MIPS~24~$\mu$m bands. We download the data created by the DustPedia consortium \citep{Davies2017, Clark2018}\footnote{\url{http://dustpedia.astro.noa.gr/}}.
The IRAC map is used in Sect.~\ref{sec:miriVP}. That image shows a bright artefact covering most of the area around the disk. To reduce its impact, we remove a background after cutting out pixels with radial distance larger than $\pm 7$~kpc from the centre (see Sect.~\ref{sec:miriVP}). Unfortunately, this does not completely remove the artefact.
The MIPS map was converted from surface brightness to star-formation-rate surface density using the prescription in \citet{Leroy2019}.

\paragraph{Optical data ---}
We use $V$-band data from \citet{HowkSavage1999}, taken at Kitt Peak National Observatory using the WIYN telescope, along with NIRCam data, to create a three-colour image (Fig.~\ref{fig:RGB}).
We also downloaded from the Isaac Newton Group Archive observations taken in the $BRI$ filters with the Wide Field Camera at the Isaac Newton Telescope (program i07bn001, PI: Trager); we processed them using standard data reduction methods and used them to build emission vertical profiles (Sect.~\ref{sec:nircam}).
These maps are in ``photometric densities'' units and we do not attempt a conversion. Instead, we will normalise them all in a similar fashion, and investigate changes in the shapes of the profiles.
We also use \textit{Hubble} Space Telescope/NICMOS Pa-$\alpha$ imaging to identify regions in the disk with high star formation rate. We use data from proposal ID~7919 (PI:~Sparks) described in \citet{Boker1999}.

\paragraph{X-Ray data ---}
We briefly compare X-Ray emission with the location of bubble-like structures. To do so, we download XMM-Newton observations of NGC~891 from the archive. We use the EPIC~S002 data from the program ID 0670950101 \citep[PI:~Bregman][]{HodgesKluckBregman2013}.

\section{Studying the disk and layered CGM with NIRCam}
\label{sec:nircam}
\subsection{Resolving dust extinction and young star clusters}
\begin{figure*}
    \centering
    \includegraphics[width=\textwidth, clip, trim={0.5cm 2.25cm 2cm 2cm}]{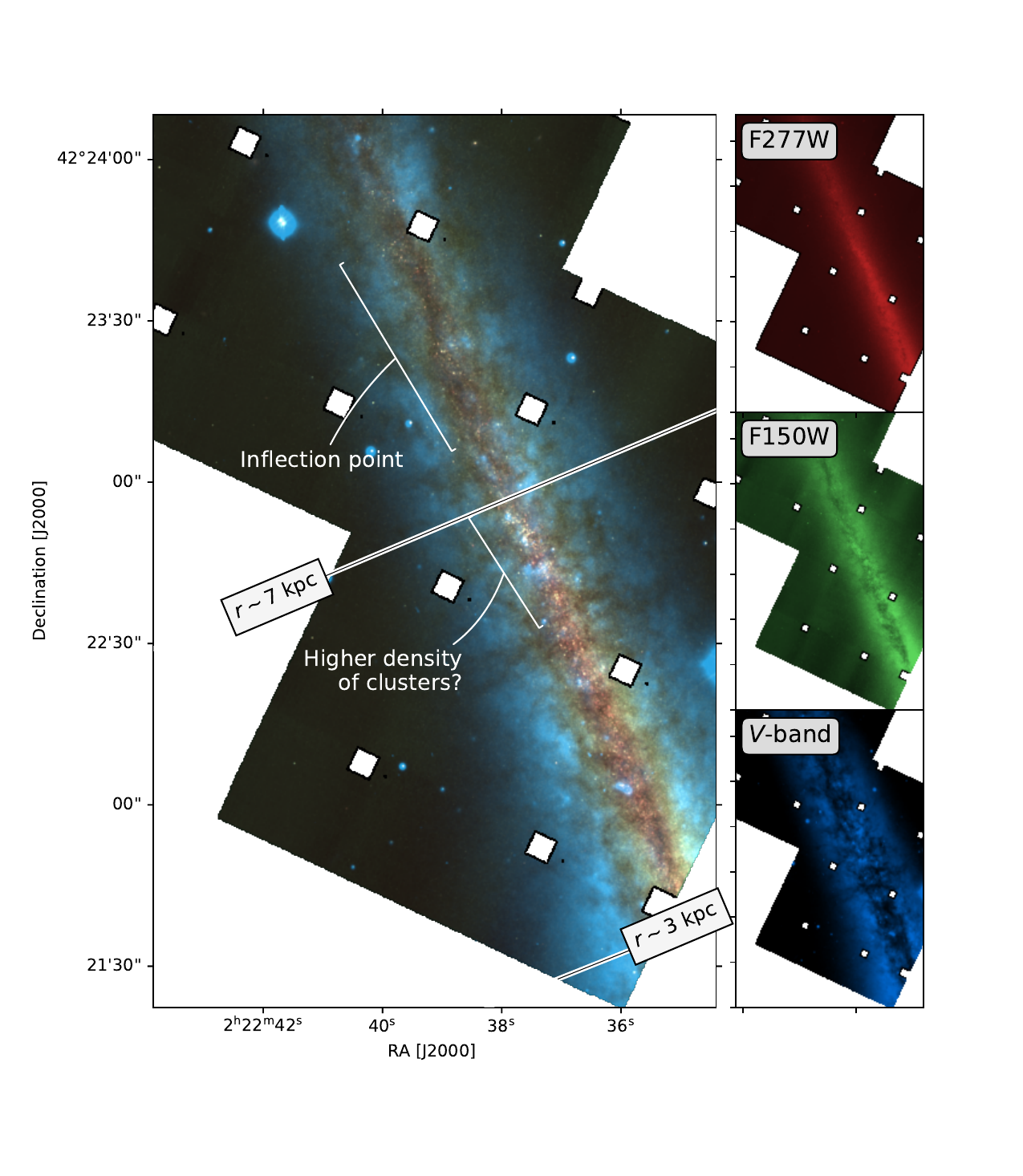}
    \caption{Multi-colour image with red: F277W, green: F150W, and blue: V-band, showing the coverage where all data overlap (the V-band and F277W images were adjusted to show the same spatial coverage, including dither gaps), and individual images shown in insets on the right. The NIRCam data was convolved to a $0.2''$ Gaussian PSF. 
    We indicate galactocentric radii $r$ at 3 and 7~kpc.
    Conspicuous dark features show dusty filaments seen in extinction perpendicular to the disk. We can also notice the presence of star clusters, some obscured by dust, and some are not.}
    \label{fig:RGB}
\end{figure*}

Figure~\ref{fig:RGB} shows a composite image of V-band, F150W, and F277W of the northern part of the disk in NGC~891. It reveals the filaments and spurs of dusty material that appear as dark lanes above the mid-plane due to the extinction of stellar light located behind the dust. These dusty obscuring structures trace the cold material, likely entrained in galactic outflows, and not infall given their dust content, and that we detect in emission with MIRI at even higher scale-heights above the disk. Similar structures were observed in recent JWST observations of M82 \citep[][Fisher et al., subm.]{bolatto2024}.

The RGB image also reveals the presence of several embedded star clusters within the thin disk, which were not visible in the V-band image due to high levels of dust obscuration, but that show up in both of the NIRCam images. These come out as bright, red-ish point sources. A separate paper will focus on these clusters.
In particular, the region with a high density of star clusters corresponds to a region with bright MIPS~24~$\mu$m emission, and has been suggested to correspond to an accumulation of \ion{H}{ii} regions along the line-of-sight \citep{Katsioli2023}, in line with the high number density of star clusters in that area. 
The star clusters are distributed within the thin disk. It is less obscured in the lower half of the image, which is consistent with the idea of trailing spiral arms in NGC\,891 with the \ion{H}{ii} regions located in front of the spiral arms on the approaching north-east side of the disk \citep{Kamphuis2007}. This also explains the detection of many embedded star clusters in this northern part of the galaxy. However, moving further away from the centre of the galaxy, the thin disk appears more obscured beyond $7-8$~kpc with maybe a smaller contribution from embedded star clusters (future work focused on NIRCam will unveil more stellar and cluster properties). At these galactocentric radii, we are likely tracing cold dense dusty clouds in the spiral arms. This is also supported by the clear warp or twist (inflection point, in Fig.~\ref{fig:RGB} that is seen in the orientation of the thin disk around radii of $\sim 8$~kpc, and is conspicuous in the F150W inset.
We do not have a clear explanation at this stage. One might be that we are tracing two different parts of a single spiral arm, or two different spiral arms, one particularly curved, and a second one on the other side of the disk. 

\subsection{Vertical disk partition}
\begin{figure}
    \centering
    \includegraphics[width=0.5\textwidth]{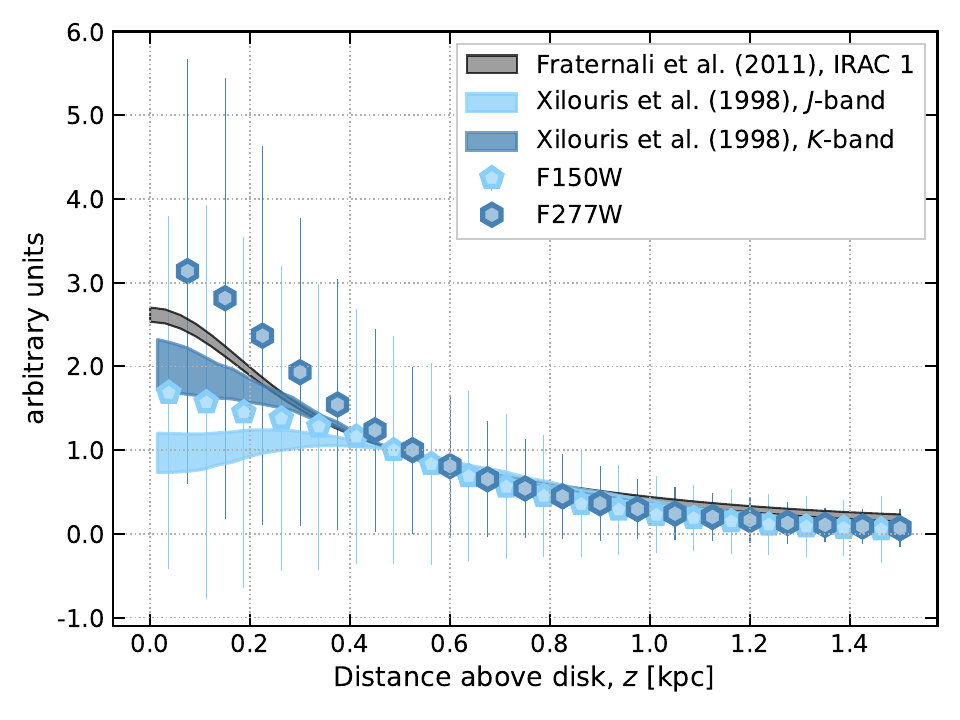}
    \includegraphics[width=0.5\textwidth]{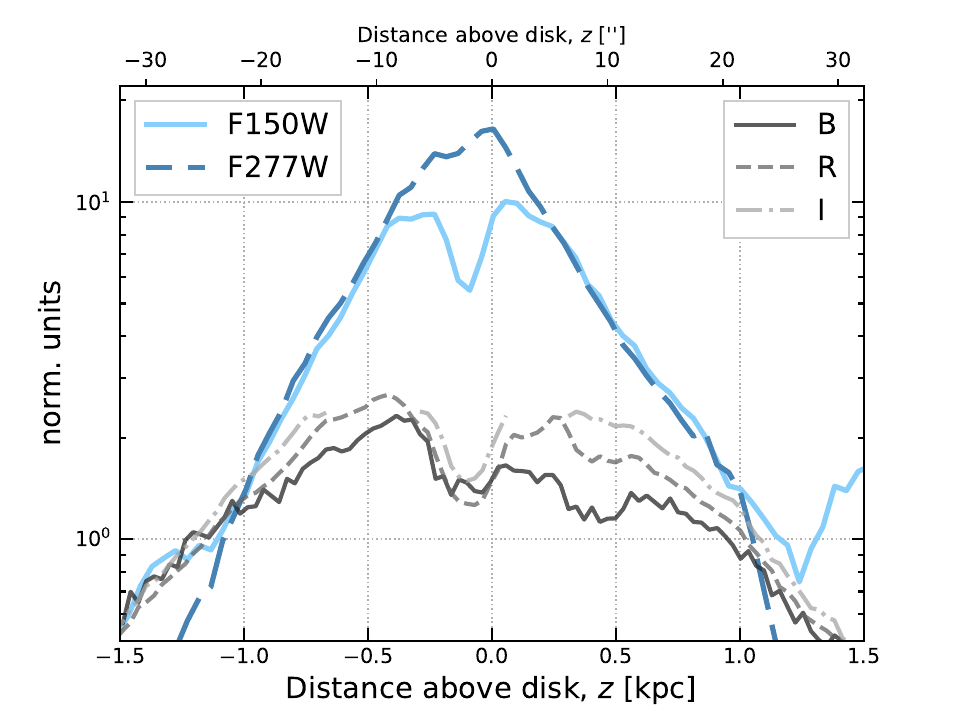}
    \caption{
    \textit{Top:} NIRCam vertical profiles shown with symbols, in light blue for F150W and dark blue for F277W. We also show the fit from \citet{Fraternali2011} using IRAC data (black line) and two fits from the radiative transfer models of \citet[][]{Xilouris1998}. In all cases, we average the vertical profiles over galactocentric radii $2.5~{\rm kpc}\leq r \leq 9~{\rm kpc}$, so that the radial coverage is the same.
    \textit{Bottom:} Emission profiles across the disk at $r \sim 8~$kpc showing the NIRCam data, and three optical bands. The images have all been normalised to the average values at $z = 1~$kpc, due to different units, for an easier comparison.
    The upturn at 1.2~kpc in F150W is an artefact due to one of the dither offsets mentioned in the text.}
    \label{fig:NIRCamVP}
\end{figure}

The available NIRCam bands offer the possibility to study the vertical extent of the stellar component. The top panel in Fig.~\ref{fig:NIRCamVP} shows the median flux (normalised at $z=0.5$~kpc) in bins of vertical distance (about 75~pc bins) for F150W (light blue) and F277W (dark blue), averaged above and below the disk, for radii between 2.5 and 9~kpc (to match the radial coverage between all profiles shown). The error bars represent the scatter in each bin (rather large, given the remaining data reduction caveats).

Several works have derived the scale height of the stellar distribution in NGC~891 using optical and near-infrared photometry, as well as radiative transfer simulations.
This approach also led to inferring the presence of a thin \emph{and} thick disks.
For example, the work of \citet{Xilouris1998} used radiative transfer models to derive stellar scale height parameters in optical and near-IR bands, studying variations both in the radial, $r$, and vertical, $z$, directions simultaneously and including a bulge component.
In Fig.~\ref{fig:NIRCamVP}, we show their best-fit profiles in the vertical $z$ direction for the $J$- ($\sim 1.2~\mu$m) and $K$- ($\sim 2.1~\mu$m) bands, for a range of radii $2.5~{\rm kpc} \leq r \leq 9~{\rm kpc}$.
Other radiative transfer modelling by \citet{Schechtman-Rook2013} even led to suggesting the existence of an extra-thin disk.
Using \textit{Spitzer}/IRAC~3.6~$\mu$m photometry, \citet{Fraternali2011} also showed a good match to NGC~891's vertical and radial profiles using a ``bulge+two disks'' model, which led to a derived thin disk scale height of 0.25~kpc while the thick disk scale height was fixed to 0.80~kpc (note that the decomposition was done with {\sc galfit} which implies ${\rm sech^2}$ functions for the disks). 
We show their fit on the NIRCam data as a thin black line.

In our case, it is very difficult to obtain conclusive fits to the vertical profiles.
We tested several approaches, most of which led to unsatisfactory results, and are therefore not shown in this study. 
We attempt to fit simple single- or double-exponential profiles in the vertical direction only, applied to different versions of the data: at native resolution, convolved to a 1\farcs0 Gaussian, and/or rebinned to bigger pixels (see Sect.~\ref{sec:extraprocessing}).
We also tested a ${\rm sech^2}$ profile fit. 
For F150W, single- and double-exponential vertical fits yield a unique value for the scale-height, $z = 0.43$~kpc, simply repeated if using a double-exponential disk.  
In F277W, we find $z = 0.45$~kpc for a single-exponential fit and $z_1, z_2 = 0.31, 0.58$~kpc for a double-exponential fit.
Using a ${\rm sech^2}$ profile leads to scale-heights of 0.61 and 0.66~kpc in F150W and F277W, respectively. However, both these fits can easily be better adjusted by eye, meaning they are not quite ``best fits.''
We found that the large scatter (used as error) in the plane of the disk due to the visible point sources, and/or background irregularities on each side of the plane, leads to a poor fit, forcing the flux values at high altitudes to be (wrongly) more constraining. 
In none of the cases we tried are the associated errors reasonable to make them trustworthy results, and we therefore do not show them\footnote{We used the Python package {\sc scipy} and its {\tt curve\_fit} function.}. 
A separate paper focused on NIRCam and other optical/near-IR bands will tackle this issue in better details.
Differences with values of scale-heights reported in the literature may be related to modelling assumptions that are not addressed here, e.g., the use of a bulge component and the simultaneous variations in both radial and vertical directions, or taking into account absorption lanes, which has been shown to affect the estimation of scale-heights \citep{Xilouris1998, Savchenko2023}.
Given the current quality of the NIRCam data, we do not investigate these avenues further in this paper.
A separate study focused on resolving NIRCam data reduction issues and subsequent analysis will tackle the vertical stratification of this edge-on galaxy in the near-IR.

To allow for a better visualisation of the vertical structure without averaging in the radial direction, the bottom panel of Fig.~\ref{fig:NIRCamVP} shows the vertical profiles of the NIRCam data at an arbitrary galactocentric radius ($r \sim 8$~kpc), normalised to the mean value at $z \sim 1~$kpc. The dust extinction is clearly visible in the dip of the F150W emission (around $-0.1$~kpc; light blue line), while the F277W data shows only a mild feature.
We compare these profiles to optical emission in the $B$, $R$, and $I$ bands, in grey.

\section{Resolving the inner circumgalactic medium with MIRI~F770W}
\label{sec:miri}
\subsection{Distribution of circumgalactic dust}
\label{sec:miriVP}
While the presence of circumgalactic dust in NGC~891 has been known for a while \citep[e.g.,][]{HowkSavage1997, Burgdorf2007, Whaley2009}, the resolution of prior instrumentation (e.g., ISO, \textit{Spitzer}, \textit{Herschel}) did not allow a study of its intricate structure on scales smaller than 100~pc. With JWST, we are able to map out the resolved circumgalactic dust emission in a 3.3~kpc-wide strip of extra-planar material around a galactocentric radius of $r=0$~kpc extending from $z \sim 0.5$ to $\sim 10$~kpc above the disk, down to scales of 12~pc (i.e., 0\farcs25 FWHM of MIRI~F770W, see Fig.~\ref{fig:finalimages}). The high spatial resolution of JWST shows that the inner CGM material is not homogeneously distributed, but is rather dominated by small-scale structures (e.g., arcs and filaments tens of pc wide and hundreds of pc long) where densities are locally enhanced (see Fig.~\ref{fig:finalimages}). 
On the smallest scales, we observe several clumpy structures that resemble those seen in the CGM of the starburst galaxy M82, with JWST (Fisher et al., subm.).
They also somewhat mirror cool clumps have been inferred through observations of \ion{Mg}{ii} absorbers \citep[e.g.,][]{Menard2010, LanFukugita2017}, but these were found on larger scales.
These cool clumps have average hydrogen number densities of 0.1~cm$^{-3}$ and gas temperatures of $T_{\rm gas} \sim 10^{4}$~K, and were postulated to have formed through thermal instabilities and/or shock compression \citep[][]{Buie2018, Buie2020, Liang2020}. They are also the perfect sites for grain processing to take place \citep{HirashitaLan2021}. 
A possible mechanism to explain the presence of dust in these clumps would involve {\small (1)} large grains ($a >0.01-0.03~\mu$m) being expelled from the disk through galactic winds \citep[][]{Zu2011, Aoyama2018, Schneider2020, Richie2024} and radiative feedback processes \citep{Sivkova2021}, and {\small (2)} subsequently undergoing shattering, transferring the predominantly large grain population to the smaller grain sizes that are detected in the JWST/MIRI F770W filter.
That said, if radiation pressure dominates, it should not be dependent on size distribution and small grains could directly be pushed out of the disk, too.
We will discuss the statistical properties of these small-scale clump structures in an upcoming paper (Chastenet et al., in prep).

\begin{figure}
    \centering
    \includegraphics[width=0.5\textwidth]{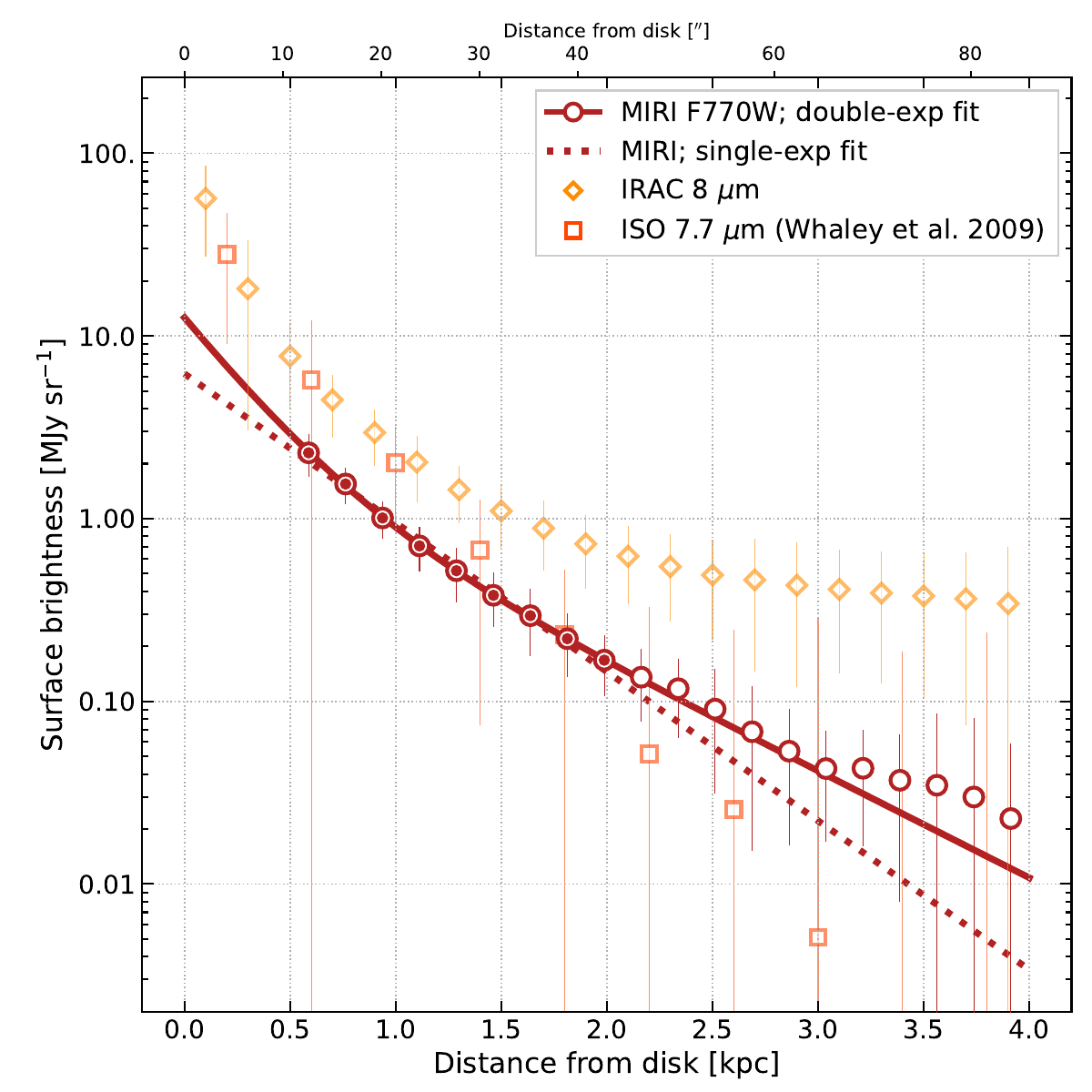}
    \caption{Vertical profiles of the MIRI~F770W image at native resolution (red; the MIRI measurements do not extend inward of 0.5~kpc, Fig.~\ref{fig:finalimages}), \textit{Spitzer}/IRAC~8~$\mu$m (shades of orange), and ISO~7.7~$\mu$m (dark-orange), with error bars showing one standard deviation in each bin.
    The filled symbols mark the ones used for fitting a double-exponential (unless mentioned otherwise) profile to the running medians, which must have ${\rm S/N \geq 3}$. See text for details on the fits.}
    \label{fig:MIRIradprof}
\end{figure}

Figure~\ref{fig:MIRIradprof} shows the vertical profiles of mid-infrared observations. 
We compute the median fluxes in the MIRI~F770W map in bins of distances and fit a double-exponential profile to this distribution, following
\begin{equation}
    S_\lambda(z) = S_{\rm d,1} \times {\rm exp} \left ( -\frac{z}{z_{\rm 1}}\right ) + S_{\rm d,2} \times {\rm exp} \left ( -\frac{z}{z_{\rm 2}}\right )\;,
\end{equation}
where $S_\lambda$ is the measured surface brightness in \mjysr, $z_1, z_2$ are the dust scale-heights, and $z$ is the distance from the disk. In Fig.~\ref{fig:MIRIradprof}, the filled symbols are those used for these fits that pass a 3$\sigma$ S/N cut, and only for $z \leq 2~{\rm kpc}$.
The red lines show the fit to the MIRI~F770W data, with the dashed one being a fit with a single-exponential, and the solid-line with two; the latter leads to scale-heights of $(0.29\pm0.65, 0.75\pm0.78)$~kpc, where the large uncertainties are due to lack of coverage all the way to the disk.
For reference, we show IRAC and ISO data. In diamond symbols, we show the binned medians of the IRAC~8 image cutting out pixels further away than $\pm 7$~kpc \textit{in radius}, similarly to what was done in \citet{Bocchio2016}. 
The square symbols show the binned medians in the ISO~7.7~$\mu$m map used by \citet{Whaley2009}.
There is an apparent offset between these ancillary data and the MIRI profile: we point out that there is a clear, bright artefact in the IRAC~8 image, covering the area we are interested in. Our efforts have not managed to bring it down to a reasonable, background-like level. Additionally, both the IRAC and ISO data are at a (much) coarser resolution, with very different transmission curves, which will inevitably affect the measured fluxes. Finally, the spatial coverage of these three (MIRI, IRAC, ISO) datasets is quite different, and sample a large range of (non-overlapping) radii and altitudes.
However, the IRAC profile does show a very similar shape to that of the MIRI data.

The scale-height of the thick dust disk measured from the MIRI~7.7~$\mu$m map is consistent with values from the literature around those wavelengths \citep[$\sim 0.3-0.6$~kpc]{Whaley2009, Bocchio2016}, and suggest a steep decline in the abundance of the smallest dust grains going away from the galaxy mid-plane further out into the inner CGM. Given the higher scale-heights inferred for large grains from far-IR and sub-mm imaging (thick disk: $z_2 \sim 1$~kpc at 250~$\mu$m, \citealt{Bocchio2016}), this suggests that the abundance of small grains decreases more steeply than the mass fraction of large grains with altitude above the mid-plane (although note that small and large grains will be sensitive to different illuminating radiation fields). This result is at odds with results in the literature, including for NGC~891 \citep{Katsioli2023} and will require further analysis to understand how the small-to-large grain ratio is altered in the disk-halo connection region of NGC~891 and other galaxies. 
Note that the distance threshold at which to stop considering data for exponential fit (here, 2 kpc) has a significant impact on the fit. This is most likely due to the lack of data all the way to the galaxy mid-plane, in the case of the MIRI data. Considering that the largest scale-height is smaller than 1 kpc, the lack of coverage below that vertical height becomes critical.

\subsection{Super-bubbles in NGC~891}
\begin{figure*}
    \centering
    \includegraphics[width=\textwidth, clip, trim={0 0 0 3cm}]{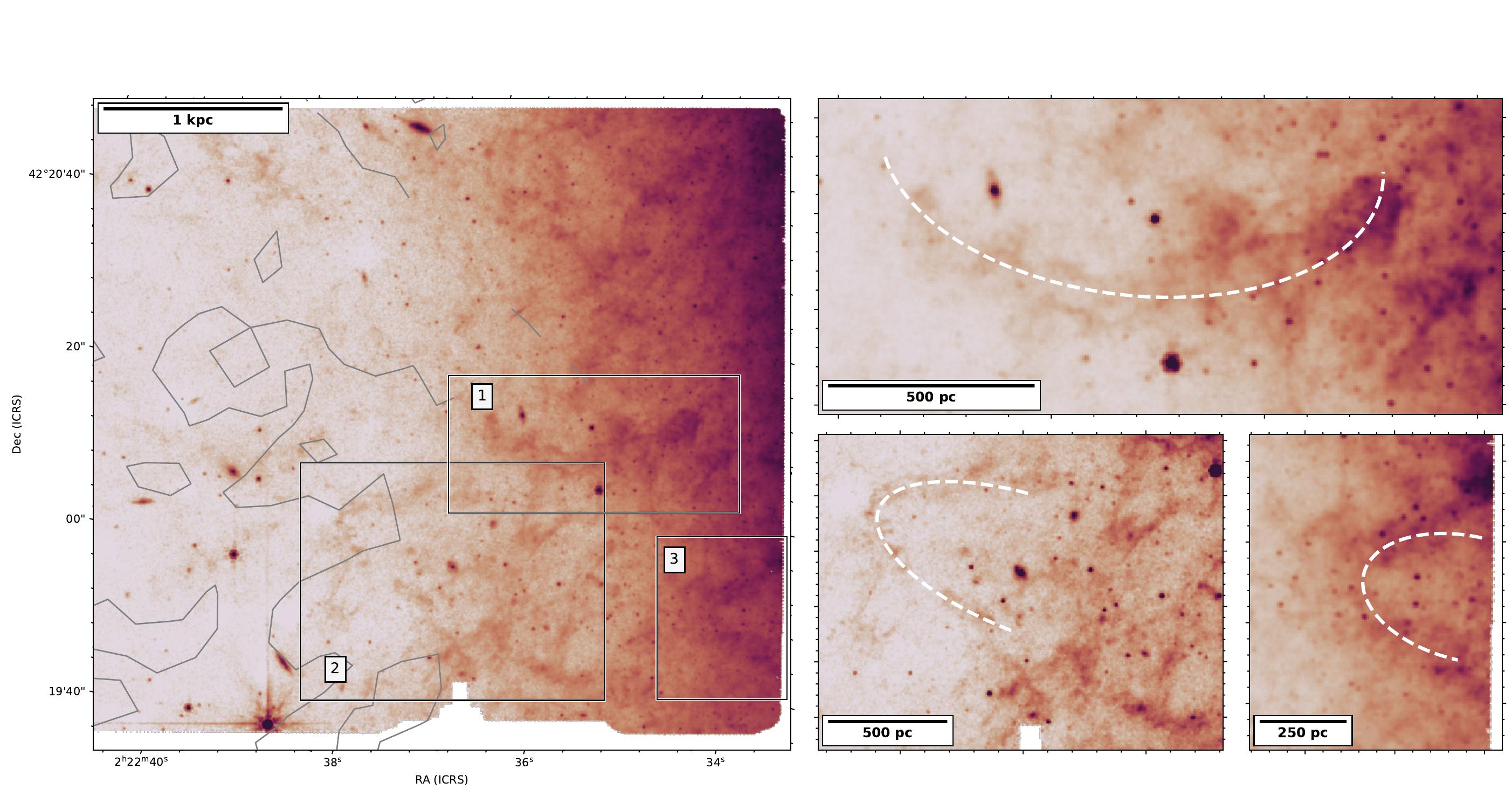}
    \caption{Cut-out of the MIRI strip closest to the disk in the left panel, with soft X-ray 95$^{\rm th}$ percentile contours in grey. We bring out three locations showing visible extended structures in boxes 1 through 3. These are shown on the right hand side, marked with arc-likes structures (by hand).}
    \label{fig:arcs}
\end{figure*}

On scales of a few tens of parsecs, we are able to discern several prominent outflow features above the disk including dusty chimneys, arcs, and parabolic structures that resemble large-scale bubbles. Some of these structures are shown in the MIRI strip, and in zoomed insets in Fig. \ref{fig:arcs}. The first panel (1) reveals an arc-like structure that resembles the walls of a former super-bubble after breaking out of the thin disk. The shells in panels (2) and (3) might correspond to similar super-bubbles where multiple SNe in a star cluster were needed to create a structure of this size \citep[see, e.g.,][for a discussion of pre-supernova feedback mechanisms with \ion{H}{ii}~region sizes]{Barnes2022}. These bubbles are formed through the continuous energy input from SNe and stellar feedback processes, and expand into the surrounding ISM creating a shell of swept up interstellar material. While the swept-up material collapses into a cool dense shell \citep[][]{Castor1975}, the interior of the bubble contains tenuous, hot ($>10^{6}$~K) gas that should be visible in low-energy X-ray emission. Matching the location of these dusty super-bubbles with XMM-Newton soft X-ray (0.3–2.0 keV) observations \citep{Temple2005}, we can see that the shell structure in panel (3) is filled with low-energy X-ray radiation. This confirms our hypothesis that these dense shell correspond to the walls of super-bubbles. Due to sensitivity limitations, we are unable to verify whether the other bubbles further out are similarly filled with a hot tenuous gas.
Examples of such super-bubbles have been observed in our own Milky~Way in the Cygnus super-bubble and Aquila super-shell with diameter sizes of $0.5-1$~kpc \citep[][]{Cash1980, Maciejewski1996}. 
Similar shells were already detected through dust extinction in the optical and H$\alpha$ emission in NGC~891 with similar super-bubble sizes up to 1~kpc \citep[e.g.,][]{Rossa2004}.
The sizes observed here are larger than those found in a recent study by \citet[][]{Watkins2023}. However, these bubbles are found in a sample of low-inclination galaxies, and at different radii, which leads to finding different morphologies and sizes due to the varying galaxy properties.

Within the dense ISM of the thin disk, the shell expansion will decelerate until the super-bubble breaks out of the thin disk and can accelerate again in the direction of the pressure gradient due to the exponential fall off in density---creating a negative density gradient---above the disk. When the super-bubble accelerates into the halo, the interface between the hot bubble interior and the dense cold shell becomes prone to instabilities such as Rayleigh-Taylor instabilities creating finger-like structures. Eventually, turbulent mixing of the two layers will cause the shell to break up \citep[][]{Baumgartner2013}, which often leaves structures that resemble dusty chimneys as observed in NGC~891 \citep[e.g.,][]{HowkSavage2000}. Numerical simulations of the breakout of super-bubbles \citep[][]{MacLow1988, MacLow1989, Gatto2015} suggest that the super-bubble could survive out to 1 to 2 vertical scale heights before the cold dense shell starts accelerating and will start fragmenting. 

The top of the super-bubble structure in Panel (3) of Fig.~\ref{fig:arcs} is located about $z=1.3$~kpc above the mid plane, while the super-bubble in Panel (2) extends out to $z=3.2$~kpc above the disk. These bubble have similar sizes compared to the scale heights of the thin and thick disks in NGC~891. Specifically, the scale heights of the thin and thick dust disk were estimated to be around 0.24 and 1.40~kpc, respectively, based on the \textit{Herschel}/SPIRE 250~$\mu$m images \citep[][]{Bocchio2016}. The dust scale height of the thick stellar disk is similar \citep[1.44~kpc;][]{Ibata2009}, while the \ion{H}{i} gas has a scale height of $\sim2.6$~kpc \citep[][]{Oosterloo2007} and the warm ionised gas can reach scale heights of 4.5~kpc \citep[][]{Dettmar1990}. At these scale heights, one can expect to soon start seeing a fragmentation of the super-bubble shell, whereas the shells currently still look intact.  \citet{Yoon2021} have studied a dusty super-bubble in NGC~891 with a diameter size of 8~kpc extending out to at least 7.7~kpc vertically upwards from the mid-plane. This super-bubble was also detected in X-rays \citep[][]{HodgesKluck2012, HodgesKluck2018}. \citet{Yoon2021} suggested that this super-bubble could develop into a galactic wind before breaking apart and returning to the disk of NGC~891 in the form of galactic fountains. The super-bubbles in this work are detected at lower altitudes above the disk, and it is unclear whether some of these bubbles can develop into a galactic wind in the future, or will rain back down on the galaxy disk after substantial cooling.

The detection of several such super-bubble structures confirms that stellar feedback processes play an important role in driving galactic outflows, even in galaxies with moderate SFR ($\sim 5~{\rm M}_{\odot}$~yr$^{-1}$) and SFR densities ($\sim 0.03~{\rm M}_{\odot}$~yr$^{-1}$~kpc$^{-2}$) like NGC~891 \citep{Yoon2021}. While NGC~891 is considered a Milky~Way-type galaxy, its prominent CGM component suggests that star formation activity and the mass loss history may have been more violent in the past (although empirical evidence may lack).

\subsection{Connecting extra-planar dusty filaments to the disk}
The MIRI images shows many fine structures within 2~kpc from the mid-plane. Unfortunately, the MIRI data do not cover the disk. 
As several studies have identified filaments coming from the disks of galaxies and extending in the CGM \citep[e.g.,][]{Rand1990, HowkSavage1997, Irwin2007}, we attempt to connect the visible structures in the MIRI/F770W map to the disk of NGC~891 using other data sets. 

We use the \textit{Spitzer}/IRAC~8~$\mu$m imaging, and apply to it the unsharp masking technique (this approach was similarly used in, e.g., \citeauthor{Yoon2021} \citeyear{Yoon2021}).
The IRAC image is blurred using a median filter box of 10 pixels in size, and that blurred image is removed from the original. The unsharp-mask IRAC image enhances very clearly filaments coming out of the disk. 

\begin{figure}
    \centering
    \includegraphics[width=0.5\textwidth, clip, trim={0 1.25cm 0 2.75cm}]{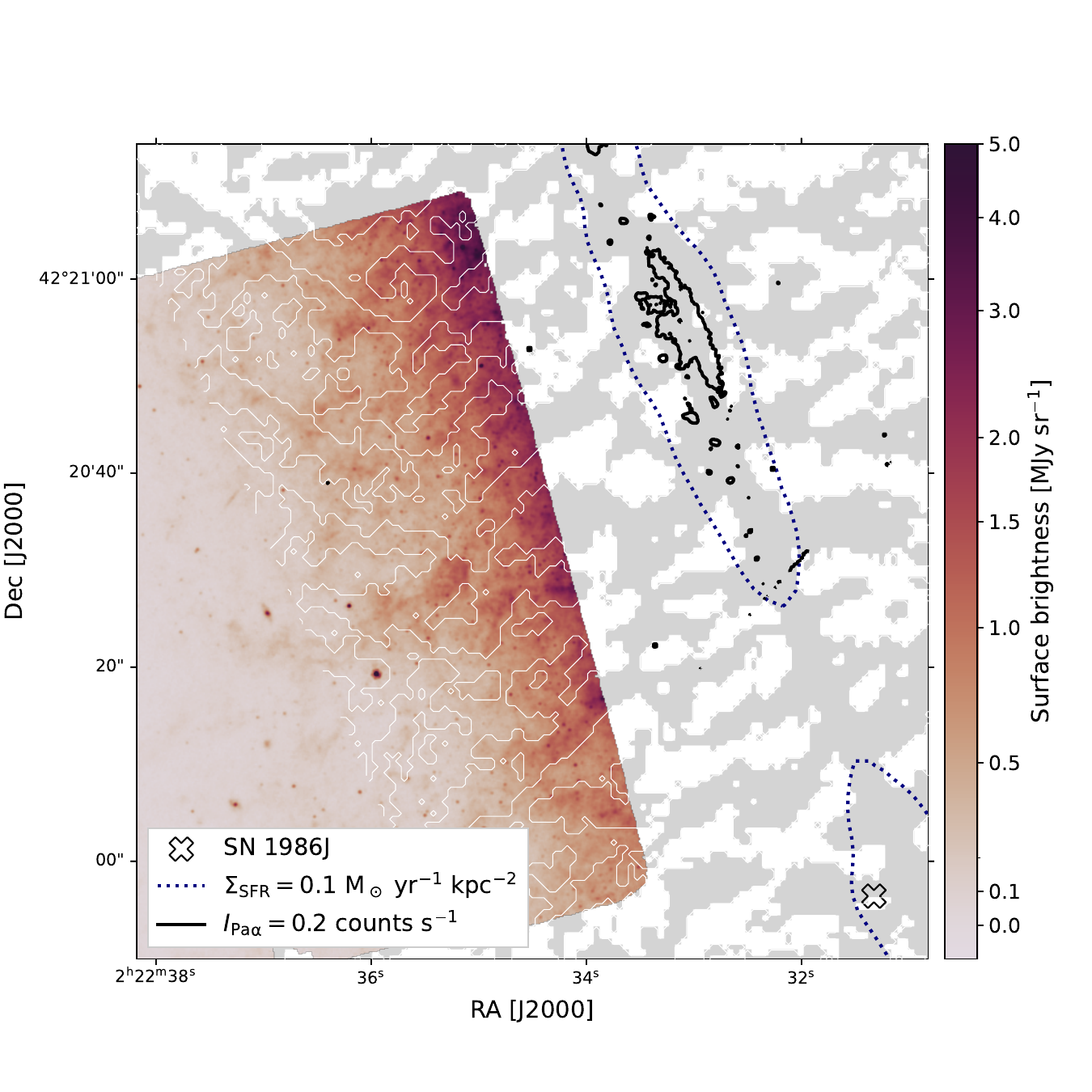}
    \caption{IRAC~8~$\mu$m unsharp-mask, shown in grey, with the MIRI image overlayed. The white contours show the continuation of the structures in the IRAC unsharp-mask, nicely following some of the structure visible in high-resolution in the MIRI map.
    We show contours of star formation rate surface density of 0.1~M$_\odot$~yr$^{-1}$~kpc$^{-2}$ derived from the MIPS-24~$\mu$m image, and intensity of Pa-$\alpha$, another tracer of recent star formation.
    The cross at the bottom-right shows the location of SN1986J.
    }
    \label{fig:FilMatching}
\end{figure}

In Fig.~\ref{fig:FilMatching}, the grey mask shows the IRAC~8 with an unsharp-mask applied. We show the MIRI image at native resolution, with white contours propagating the unsharp-mask filaments. By eye, the agreement seems rather good, with darker colour (higher flux density) often falling within the IRAC~8 filaments\footnote{Note that we could qualitatively observe the same dusty filaments connecting back to the disk in Fig.~\ref{fig:RGB}.}.
Some of these overlapping pixels can be traced back to the disk, where we also show in black contours the Pa$\alpha$ intensity, as a tracer of recent star formation, as well as $\Sigma_{\rm SFR}=0.1$~M$_\odot$~yr$^{-1}$~kpc$^{-2}$ inferred from MIPS~24 \citep[][]{Leroy2019}, a threshold at which star formation can be responsible for blowing material off the mid-plane, suggested by different studies \citep[e.g.,][]{Heckman2002, MMT2011}.

\subsection{Origin of the MIRI/F770W emission}
The detection of emission originating from PAHs out to $\sim 4$~kpc is remarkable given the hostile environment that a galaxy halo represents with its X-ray emitting diffuse ionised gas component, in which the smallest dust grains could be destroyed almost instantly through sputtering processes \citep{Micelotta2010}. That said, NGC~891 is not a standalone case. For example, PAH emission was detected out to 6~kpc in M82 (\citealt{Engelbracht2006}; see also follow-up work with JWST by \citealt{bolatto2024} and Fisher et al., subm.). 
Such extended emission from PAHs was also found in less extreme galaxies: NGC~5907 \citep{IrwinMadden2006}, NGC~5529 \citep{Irwin2007}, and NGC~5775 \citep{Lee2001}. 

Since the claimed detection originates from broad-band filters, the observed emission could be contaminated by other emission sources, and not simply trace the mid-IR feature. Beyond the thick stellar disk in NGC~891 ($h_{\text{z}}\sim 400-800$~pc), we can safely assume that the stellar emission will be only minimally contributing to the MIRI~F770W flux at high latitudes. Recent MIRI~MRS observations of four pointings at 0.5 and 1~kpc above the disk (from the same Cycle~1 GO~\#2180) also suggest that strong contributions from ionised gas lines are unlikely in that filter (Chastenet, De~Looze et al., in prep.).
The same MRS spectra do detect carbon emission features, similarly to \cite{Rand2008}, who spectroscopically identified mid-IR emission features above the disk of NGC~891 at $\lambda > 10~\mu$m.
This makes us fairly confident that the MIRI~7.7~$\mu$m emission is originating from dust. While the MIRI~F770W filter is centred on the 7.7~$\mu$m feature from ionised PAHs \citep[e.g.,][]{Maragkoudakis2022}, there may be underlying continuum emission from very small grains ($a \lesssim 0.01$~$\mu$m). 
In dust emission models, this continuum is expected to have a non-negligible contribution to the F770W filter only when reaching very high radiation fields \citep[][]{Draine2021, Sutter2024}, or unexpectedly high fraction of small grains over large grains. 

How PAHs were able to reach such high latitudes above the disk is not yet entirely clear. Dusty material could originate from the mid-plane and 
{\small (1)} may have left the galaxy disk entrained in galactic outflows \citep{Kannan2021, Richie2024},
{\small (2)} was pushed out through radiation pressure (Sivkova et al. in prep), 
{\small (3)} was transported by cosmic ray-driven winds, which seem to be slower and cooler, and therefore smoothly lift material away from the disk \citep[][]{Girichidis2016CR, Girichidis2018CR, Girichidis2022CR, DeFilippis2024}.
The second option is that dense cloudlets in the CGM are created through condensation of CGM material via thermal instabilities in the halo \citep[][]{Buie2018, Liang2020}, but these structures are not expected to host any PAHs originating from the disk. They may potentially harbour PAHs that formed in-situ through AGB stellar mass loss \citep{Groenewegen1997, Mauron2008}, although it is unlikely that this will significantly contribute to the PAH abundance in the halo since these freshly formed grains would be ejected in the hot X-ray emitting halo gas and would not stand a chance to survive for long \citep[e.g.,][]{Scannapieco2017}.

Given that the dusty filaments connect back to star-forming regions within the disk, we consider it most likely that the dusty material traced with our JWST observations at altitudes of $\sim 1 -4$~kpc above the disk have been removed from the galaxy disk through galactic winds. In that case, the dusty material of the disk should still be present in these structures and the JWST observations give us a unique view on how material has been slung out the galaxy disk and how PAHs are processed in the harsh environment of the CGM. In the next paragraphs, we discuss what physical conditions are needed for PAHs to survive out to these large extraplanar distances. 

\subsection{Survival of PAHs out to 4~kpc}
Assuming that the extra-planar emission originates from small carbonaceous dust, we can ask ourselves how the smallest dust particles can survive out to 4~kpc above the disk. Typical outflow velocities of $50-75$~km~s$^{-1}$ for cold material leaving the disk have been inferred based on \ion{H}{i} data of NGC~891 \citep{FraternaliBinney2006, FraternaliBinney2008}, and are consistent with velocities estimated from the combined effects of radiation pressure on dust grains, gravity, and gas drag \citep[][and in prep]{Sivkova2018}. This would imply that it takes $50-80$~Myr for the material to reach these altitudes above the disk. 
We note that higher velocities are plausible, and the currently measured outflow velocities may furthermore not resemble those that pushed out material out of the disk in the past.
For example, the turbulent mixing of hot gas with cool clouds can boost the velocities of the cool material entrained in hot galactic winds. 
Also, if NGC~891 experienced a starburst episode before, the star formation feedback may have been more violent resulting in higher outflow velocities. This implies that our estimated lifetimes ($50-80$~Myr) for PAHs detected at 4~kpc above the disk are likely to be considered as strict upper limits. 
These dynamical timescales are several orders of magnitude longer than the average lifetime of PAHs in a tenuous hot gas \citep[of the order of 1000 years;][]{Micelotta2010}. 
This requires that the PAHs must be shielded from harsh environments in pockets of cool dense regions---manifested as clumpy structures and cool dense shells around the super-bubbles. PAHs may also be found in a mixing layer of cool, diffuse ionised gas that connects the warm gas and the cold dense cloudlets, providing an additional source of emission. Past works have already shown the existence of PAHs on the surfaces of cold clouds in hot gas for long periods \citep{Micelotta2010}.

Simulations \citep{SchneiderRobertson2017, TanFielding2024} demonstrate that cold ($100 - 1000$~K), dense, presumably dust-rich clouds can be accelerated out of the disk via a variety of mechanisms, including, e.g., ram pressure from super bubbles.
As these clouds experience hydrodynamic instabilities from the wind, they are mixed with the hot phase and accelerated, leading to a multiphase structure, including cold dense clumps closer to the disk, a cool (10$^4$ K) ionised phase that is being accelerated, and a transient mixed phase ($10^5$~K) that either gets incorporated into the hot phase of the wind or cools back onto the clouds.
In this scenario, the cool, $10^4$~K layer would still be able to shield small dust grains, but we would expect shattering to be more efficient in there than inside a colder cloud.
This is where potentially large grains would be destroyed to smaller size grains, continuously replenishing the small grain distribution. In addition to seeing PAH emission there, we would still observe, on a smaller timescale, emission in the hotter $10^5$~K phase, where the smallest grains would be accelerated.

The fact that PAH emission may originate from those cool dense clouds is supported by the perfect agreement between the PAH emission and optical dust extinction at scale-heights of $1-2$~kpc above the disk \citep[][Xilouris et al., in prep]{HowkSavage2000}, which would suggest that the PAH emission originates from cool regions within these winds.
Although \citet{HowkSavage2000} argue that they should be sensitive to detect dust in extinction out to 3~kpc in their deep data, they only detect dust extinction around $1-2$~kpc above the disk. While we detect emission out to 4~kpc in our MIRI~F770W map, the lack of corresponding optical dust extinction could suggest changes in the dust grain population making it less efficient in obscuring optical light (although geometry consideration might arise). 
Since grains absorb and scatter light most efficiently at wavelengths similar to their size ($\lambda$ = $2\pi$a), this would suggest that the abundance of large grains ($a\gtrsim0.1~\mu$m) drops beyond 2~kpc. This is consistent with the picture that shattering efficiently converts large grains into small grains in the circumgalactic medium \citep[e.g.,][Sivkova et al., in prep]{HirashitaLan2021}, although somewhat inconsistent with scale-heights of mid- and far-IR, as mentioned in Sect.~\ref{sec:miriVP}. 
Considering that shattering is not as efficient within the cold (100~K) phase, this could be an indication that dust processing occurs in the cool/mixed gas phase.

The relative motion of the hot wind and the cool cloud gives rise to such a turbulent radiative mixing layer. At this interface, a gas-phase with intermediate temperatures ($\sim 10^{4}-10^{5}$~K) and densities ($\sim 1~$cm$^{-3}$) can form due to hydrodynamic instabilities (mostly Kelvin-Helmholtz instabilities) and short cooling times. If cooling times are short enough, or equivalently, if clouds are large enough, the cold clouds can grow in mass and prolong their lifetime in the CGM \citep[][]{Cooper2009, GronkeOh2018, Schneider2020,FieldingBryan2022, Abruzzo2022, SchneiderMao2024}. 
It is possible that the PAH emission originates from this mixing layer, where PAHs should be able to survive for $\sim 2.5$~Myr \citep[][though their calculations are for spherical grains]{Richie2024}, during which they can be prone to sputtering. The emission observed with JWST may correspond to PAH emission in this mixing layer that is continuously being replenished by material from the cool cloud \citep{Micelotta2010}.

Regardless of the exact process, these pockets of cold material and the gaseous material in the mixing layer would need to be shielded from the harsh radiation field in the CGM. The DIG ionisation levels suggest that UV photons leaking from \ion{H}{ii} regions in the disk, together with hardening of the radiation field, are responsible for the extraplanar DIG ionisation 
\citep[][]{Jo2018, Vandenbroucke2018, Belfiore2022, McCallum2024, McClymont2024}. A population of hot evolved stars may also impact the DIG ionisation \citep{Rand2011, FloresFajardo2011}, though \citet{Levy2019} find that these stars only contribute up to 10\% of the extraplanar DIG ionisation in normal star-forming edge-on galaxies. Since dense cloudlets in the CGM are significantly denser than their surroundings compared to the average density ($10^{-4}-10^{-2}$~cm$^{-3}$) in the hot phase, we do not expect UV photolysis to play an important role in the destruction of PAHs. \citet{Micelotta2011} argue that the UV radiation escaping from the disk is responsible for the excitation of these PAHs, but likely does not contribute significantly to the photodestruction of PAHs. The UV radiation field will be significantly stronger closer to the disk, which would imply that the smallest PAHs ($<50$ C atoms) and least stable PAHs have likely been destroyed already when the material left the disk, and the PAHs remaining in the denser structures of these galactic winds are able to withstand the UV radiation escaping to these high altitudes above the disk.

In that case, \citet{Micelotta2011} argue that the dominant mechanism destroying PAHs around 3~kpc above the disk is processing through cosmic rays. Their calculations for NGC~891 predict PAH lifetimes ranging between 100 and 1000~Myr for small ($N_{\text{C}}=60$) and large ($N_{\text{C}}=500$) PAHs, respectively. 
These timescales exceed the estimated travel time of PAHs to reach latitudes of 3~kpc above the disk, suggesting that PAHs can freely travel out to these high latitudes without being destroyed, when shielded from the harsh radiation field in dense cloudlets. Hence, the lifetime of PAHs in the CGM of NGC~891 will largely be constrained by the lifetime of the dense cloudlets hosting PAHs. The PAH emission provides a unique vantage point to study the cold dense clumps that are ejected from the galaxy disk entrained in galactic outflows. The distribution, properties and lifetime of these cloudlets will be studied in a future work (Chastenet et al., in prep). Looking at the relative abundance of ionised and neutral PAHs with the MIRI MRS spectra will allow us to confirm whether the PAH emission is mostly originating from a less dense turbulent mixing layer or from the surfaces of the cool cloudlets. Together with a careful analysis of the PAH size distribution, the MIRI MRS spectra will be crucial to understand what role UV radiation plays in the excitation and destruction of PAHs at these high distances above the disk.

\section{Conclusions}
\label{sec:conclusion}
This paper presents the first set of data from the JWST Cycle~1 GO~\#2180 program (PI Ilse De~Looze), focused on the edge-on galaxy NGC~891, and the cycling of baryonic material from the galaxy disk to the circumgalactic medium (CGM) and back.

This first paper describes the NIRCam F150W and F277W data, which observed the northern half of the disk, and the MIRI~F770W observations of a long $\sim 3 \times 10~$kpc strip of the CGM (Figs.~\ref{fig:propPatterns} and \ref{fig:finalimages}).

Combined with optical V-band observations tracing dust extinction, we find that the near-IR emission clearly highlights vertical dust plumes coming off the mid-plane that match the optical extinction lanes (Fig.~\ref{fig:RGB}). 
We compare vertical profiles of the NIRCam data with $BRI$ imaging and point out the conspicuous absorption lane visible in NIRCam~F150W, and a much milder one in the F277W filter.
The new near-IR observations reveal several embedded star clusters, compared to past optical data.

In the MIRI strip, we detect emission in the F770W filter confidently ($\rm S/N > 3$) out to $\sim 3$~kpc and can identify structures out to 4~kpc ($\rm S/N > 1.5$). 
The high spatial resolution and sensitivity of MIRI allows the identification of these fine structures in the forms of arcs and filaments (Figs.~\ref{fig:arcs}). 
Using ancillary data covering the disk of NGC~891, we can confirm that some of these filaments connect back to the mid-plane, to regions of high star formation rate (Fig.~\ref{fig:FilMatching}), suggesting that feedback-driven galactic winds play an important role in regulating baryonic cycling. 
In addition to the dusty super-bubble detected by \cite{Yoon2021} at $\sim 7$~kpc, the MIRI data hint at the presence of two other super-bubbles that extend out to $\sim 1.25$ and $\sim 3.5$~kpc.
The presence of dust in the forms of small dust grains (traced by the 7.7~$\mu$m emission), and possibly PAHs, is a clue to the destruction processes impacting these small grains and the mechanisms responsible for their ejection off the galaxy disk. We discuss several scenarios to explain the survival of dusty material at these distances, despite the somewhat harsh environment that is the disk-halo interface.
These small grains could be present in pockets of dense material and protected from ionising radiation; this scenario agrees with simulations and a good match between emission and extinction in literature; 
The emission can also come from the surface layers of clouds, where the differential wind speed between hot and warm gas is enough to create a mixing layer, replenished by cooling material from the hot gas phase, as seen in recent simulations.
A large amount of small (and suspiciously circular) clumps are clearly visible in the MIRI~F770W image, potentially supporting the hypothesis of dust transport in dense clouds. A follow-up paper will carefully identify the nature of these objects.

Additional spectroscopic data from the same program will shed more light on the shape of the mid-infrared emission, tracing nebular emission lines and carbonaceous dust features (Chastenet, De~Looze et al., in prep).
Altogether, the new data collected by JWST provide us with a unique view on how the cool disk material is transported out of the disk through galactic winds, and how long the cool pockets of material are able to survive out into the inner CGM.

\begin{acknowledgements}
We thank the referee for their careful and thorough review of the paper, which helped improved its content and clarity.
JC, IDL, and MB acknowledge funding from the Belgian Science Policy Office (BELSPO) through the PRODEX project ``JWST/MIRI Science exploitation'' (C4000142239).
MR acknowledges the support from the project PID2020-114414GB-100, financed by MCIN/AEI/10.13039/501100011033. 
R.C.L. acknowledges support for this work provided by a National Science Foundation (NSF) Astronomy and Astrophysics Postdoctoral Fellowship under award AST-2102625. 
EWK acknowledges support from the Smithsonian Institution as a Submillimeter Array (SMA) Fellow and the Natural Sciences and Engineering Research Council of Canada.
SB and VC acknowledge funding from the INAF Mini Grant 2022 program `Face-to-Face with the Local Universe: ISM's Empowerment (LOCAL)'.
SCOG and RSK acknowledge funding from the Heidelberg Cluster of Excellence EXC 2181 (Project-ID 390900948) `STRUCTURES: A unifying approach to emergent phenomena in the physical world, mathematics, and complex data' supported by the German Excellence Strategy and from the European Research Council in the ERC synergy grant `ECOGAL – Understanding our Galactic ecosystem: From the disk of the Milky Way to the formation sites of stars and planets' (project ID 855130). RSK also thanks for support from the German Ministry for Economic Affairs and Climate Action in project ``MAINN'' (funding ID 50OO2206). 
KN acknowledges support from the MEXT/JSPS KAKENHI grant numbers 19H05810, 20H00180, 22K21349, 24H00002, and 24H00241, as well as that from the Kavli IPMU, World Premier Research Center Initiative (WPI). SW gratefully acknowledges the Collaborative Research Center 1601 (SFB 1601 sub-project B6) funded by the Deutsche Forschungsgemeinschaft (DFG) – grant number 500700252.

We thank Francesco Belfiore for sharing his code to create convolution kernels, available at {\tt \url{https://github.com/francbelf/jwst_kernels}}.
We are grateful to Prof. Judith Irwin for sharing the \textit{Spitzer} and \textit{ISO} data from their paper \citep{Whaley2009}.
This research has made use of 
the SIMBAD database, operated at CDS, Strasbourg, France \citep[][]{Wenger2000},
APLpy, an open-source plotting package for Python hosted at {\tt \url{http://aplpy.github.com}},
SciPy \citep{Virtanen_2020},
ds9, a tool for data visualization supported by the Chandra X-ray Science Center (CXC) and the High Energy Astrophysics Science Archive Center (HEASARC) with support from the JWST Mission office at the Space Telescope Science Institute for 3D visualization,
matplotlib, a Python library for publication quality graphics \citep{Hunter:2007},
Astropy, a community-developed core Python package for Astronomy \citep{2018AJ....156..123A, 2013A&A...558A..33A}, and NumPy \citep{harris2020array}.
\end{acknowledgements}

\section*{Data Availability}
Some/all of the data presented in this paper were obtained from the Mikulski Archive for Space Telescopes (MAST) at the Space Telescope Science Institute. The specific observations analyzed can be accessed via {\url{https://doi.org/10.17909/xmww-rx26}}. STScI is operated by the Association of Universities for Research in Astronomy, Inc., under NASA contract NAS5–26555. Support to MAST for these data is provided by the NASA Office of Space Science via grant NAG5–7584 and by other grants and contracts.
 



\bibliographystyle{aa} 
\bibliography{example} 








\label{lastpage}
\end{document}